\documentclass[aps,prb,showpacs]{revtex4}
\pdfoutput=1
\usepackage{amssymb}
\usepackage{amsmath}
\usepackage{graphicx}
\newcommand{\beq}{\begin{equation}}
\newcommand{\eeq}{\end{equation}}
\newcommand{\bea}{\begin{eqnarray}}
\newcommand{\eea}{\end{eqnarray}}
\begin{document}
\flushbottom

\title{Les Houches Notes on Graphene}

\author{Antonio H. Castro Neto}

\affiliation{Department of Physics, Boston University, 590 Commonwealth Avenue, Boston, MA 02215, U.S.A.}

\date{\today}

\begin{abstract}
Graphene research is currently one of the largest fields in condensed matter. Due to its unusual electronic spectrum with Dirac-like quasiparticles, and the fact that it is a unique example of a metallic membrane, graphene has properties that have no match in standard solid state textbooks. In these lecture notes, I discuss some of these properties that are not covered in detail in recent reviews \cite{rmp09}. We study the particular aspects of the physics/chemistry of carbon  that influence the properties of graphene; the basic features of graphene's band structure including the $\pi$ and $\sigma$ bands; the phonon spectra in free floating graphene; the effects of a substrate on the structural properties of graphene; and the effect of deformations in the propagation of electrons. The objective of these notes is not to provide an unabridged theoretical description of graphene but to point out some of the peculiar aspects of this material.   
\end{abstract}


\maketitle

\section{Introduction}

Graphene, a two-dimensional allotrope of carbon, has created a immense interest in the condensed matter community and in the media since its isolation in 2004 \cite{novo04}. On the one hand, graphene has unique properties that derive from its honeycomb-like lattice structure such as the Dirac-like spectrum (that mimics effects of matter under extreme conditions), its low dimensionality (that leads to enhanced quantum and thermal fluctuations), and its membrane-like nature (that mixes aspects of soft and hard condensed matter). On the other hand, because of the strength and specificity of its covalent bonds, graphene is one of the strongest materials in nature (albeit one of the softest), with literally none extrinsic substitutional impurities, leading to the highest electronic mobilities among metals and semiconductors \cite{rise}. Therefore, graphene is being considered for a plethora of applications that range from conducting paints, and flexible displays, to high speed electronics. In fact, it can be said that perhaps, not since the invention of the transistor out of germanium in the 1950's, a material has had this kind of impact in the solid state literature. However, unlike ordinary semiconductors such as germanium, gallium-arsenide, and silicon, graphene's unusual properties have to be understood before it can really have an impact in technological applications.

Any material has a hierarchy of energy scales that range from the atomic physics ($\sim 10$ eV), to many-body effects ($\sim 10^{-3}$ eV). To understand the behavior of a material one needs to understand how these different energy scales affect its macroscopic properties. While structural properties such as strength against strain, shear and bending may depend on the covalent bonds formed by the atoms, magnetism and superconductivity are governed by the particular way electrons interact with each other through Coulomb forces. Furthermore, while the properties of metals and semiconductors depend on the physics close to the Fermi energy (a direct consequence of Pauli's exclusion principle), the nature of the vibrational spectrum depends on the particular way ions interact among themselves and how the electrons screen these interactions. 

One of the great accomplishments of the application of quantum mechanics to the theory of metals is the understanding that while different materials can be structurally very different from each other, its long wavelength and low energy physics is essentially identical and depend on very few parameters. This so-called renormalization towards the Fermi energy \cite{shankar} is one of the greatest theoretical accomplishments of the twentieth century and is the basis of Landau's theory of the Fermi liquid \cite{baym}. In systems where the low energy physics is described by a transformation of coordinates with {\it Galilean invariance} the most significant parameter is the ``effective'' mass of the carriers that acts as to generate a {\it scale} from which is possible to compute most of the important physical quantities such as specific heat, magnetic susceptibility, electronic compressibility, and so on. The most basic difference between graphene and other materials is that its low energy physics is not Galilean invariant, but instead {\it Lorentz invariant}, just like systems in particle and high-energy physics with Dirac particles as elementary excitations \cite{pw06}. In this case, the renormalization towards the Fermi energy is different from other materials \cite{gonzalez} because, in the absence of a ``mass'' (which in graphene means the absence of a gap in the electronic spectrum), all physical quantities depend on a characteristic ``effective'' velocity that plays an analogous role as the speed of light plays in relativistic quantum mechanics \cite{note1}. However, unlike true relativistic fermionic systems \cite{baym2}, the Dirac quasiparticles in graphene still propagate with a velocity that is much smaller than the speed of light, the speed that Coulomb interactions propagate. Therefore, the Coulomb field can be considered instantaneous in first approximation, making of the electrodynamics of graphene electrons a mix between a relativistic and a non-relativist problem. Clearly this unusual situation requires a re-evaluation of the Fermi liquid theory for this material. 

In these notes, however, we are going to focus on the more basic aspects of graphene's properties. In Section \ref{chemistry} we are going to discuss the $s-p$ hybridization theory and how it leads to the basic energy scales of the graphene problem. We are going to show that the problem of hybridization is controlled by the angle between the $s-p$ hybridized orbitals. Using the Slater-Koster theory, we compute the main matrix elements of the problems in terms of this angle. In Section \ref{bands} we move from molecular orbitals to the crystal and discuss the simplest tight binding Hamiltonian that describes the {\it full} band structure, that is, that includes both $\pi$ and $\sigma$ bands. We show that even this simple band-structure reproduces quite well the results of more sophisticated {\it ab initio} methods. This particular description becomes particularly good close to the Fermi energy where the Dirac particles emerge naturally. Phonons in free floating graphene are discussed in Section \ref{phonons}. We show how the flexural modes result from the bending energy of a soft membrane and how those modes can be quantized. We show that as a result of the presence of flexural modes the linear phonon theory predicts an instability of the graphene sheet towards crumpling. The effect of a substrate on the flexural mode spectrum is discussed in Section \ref{constraint}. We show that the presence of a substrate, that breaks rotational and translational symmetry, allows for new terms in the phonon Hamiltonian that change considerably the energy dispersion of the flexural modes. We also show that within the linear theory, graphene follows the substrate in a smooth way with the characteristic length scales that are dependent on the details of the interaction with the substrate. When graphene is deformed in some way, either by bending or strain, the electronic motion is affected directly. In Section \ref{deformed} we show that at long wavelengths deformations lead to new terms in the Dirac equation that couple to the electrons as vector and scalar potentials. There are cases, however, where graphene is not slightly deformed but strongly deformed in which case the Dirac theory has to be completely reconsidered. In these notes we discuss briefly the case of a graphene scroll which results from the competition between the bending energy (that favors flat graphene) and the van der Waals interaction of graphene with itself (that wants it to have maximum area overlap). This is only one example where structural deformations can have a strong effect on many of the electronic properties of this material. Our conclusions are given in Section \ref{conclusions}.

\section{The Chemistry}
\label{chemistry}

\begin{figure}[tbh]
\centerline{\includegraphics[width=6cm, keepaspectratio]{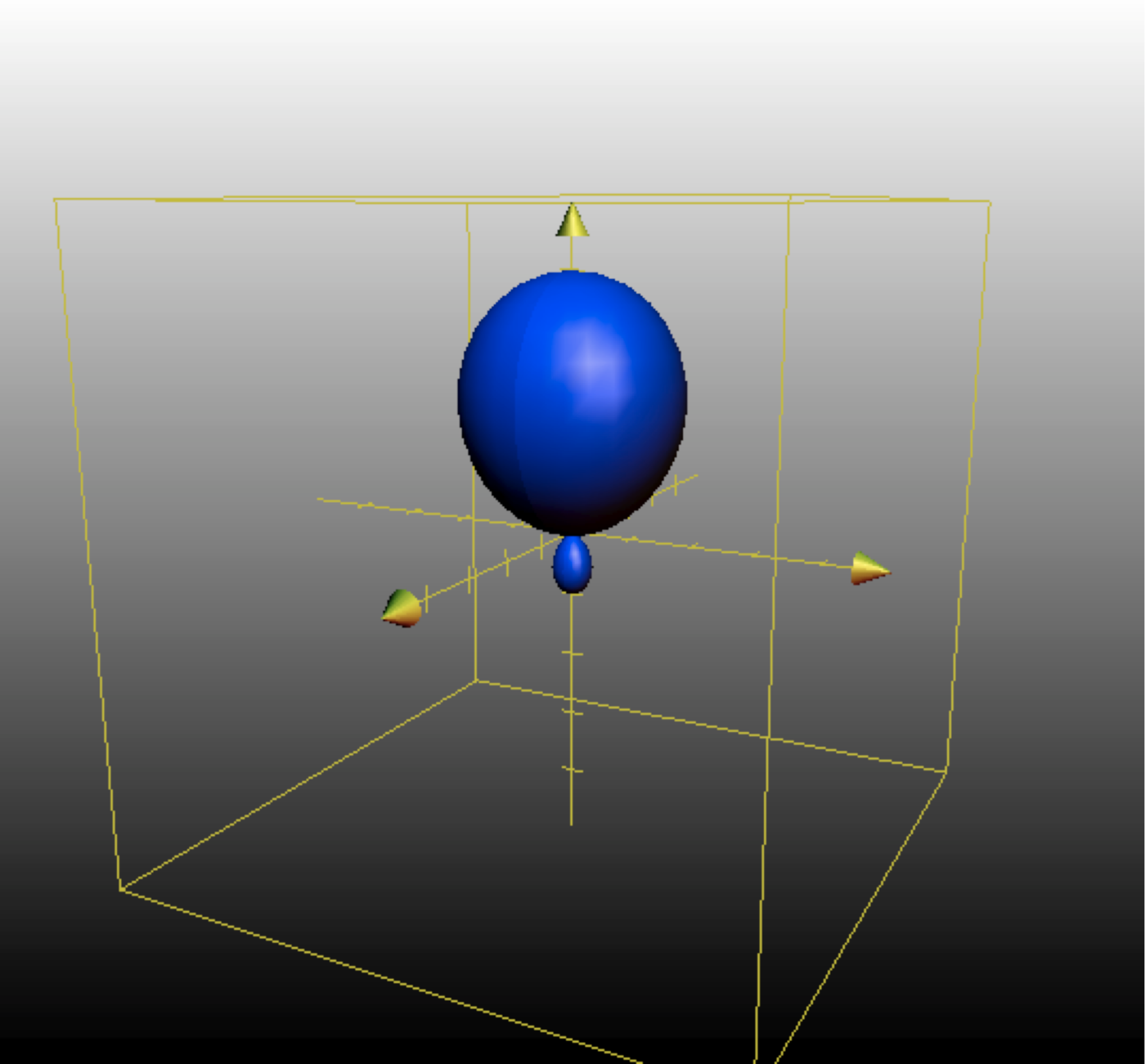}}
\caption{Absolute value of the wave-function for a hybridized $s-p$ state, Eq.(\ref{sp0}), for $A=0.5$.}
\label{spfig}
\end{figure}

The electronic configuration of atomic carbon is 1s$^2$ 2s$^2$ 2p$^2$. In a solid, however, carbon forms $s-p$ hybridized orbitals. The 1s electrons form a deep valence band and essentially all properties of carbon-based materials can be described in terms of the $2s$ and $2p_x$, $2p_y$ and $2p_z$ orbitals that can be written as \cite{Baymbook}:
\begin{eqnarray}
\langle {\bf r} | s \rangle &=& R_s(r) \times 1  \, ,
\nonumber
\\
\langle {\bf r} | p_x \rangle &=& R_p(r) \times \sqrt{3} \sin\theta \cos\phi \, ,
\nonumber
\\
\langle {\bf r} | p_y \rangle &=& R_p(r)\times \sqrt{3} \sin\theta \sin\phi \, ,
\nonumber
\\
\langle {\bf r} | p_z \rangle &=& R_p(r) \times \sqrt{3} \cos\theta \, ,
\label{sporbitals}
\end{eqnarray}
where $R_s(r) = (2-r/a_0) e^{-r/(2 a_0)}$, and $R_p(r)=(r/a_0) e^{-r/(2 a_0)}$ are the radial wave-functions. One particular way to parametrize a hybridized $s-p$ state is given by Pauling \cite{Pauling}:
\begin{eqnarray}
|0\rangle = A |s\rangle + \sqrt{1-A^2}|p_z\rangle \, ,
\label{sp0}
\end{eqnarray}
where $A$ is a parameter that describes the degree of hybridization between $s$ and $p$ states. This basic orbital is shown in Fig.\ref{spfig}. 
The energy associated with this orbital can be obtained from the hydrogen atom spectrum:
\begin{eqnarray}
E_0 = \epsilon_{\pi}=\langle 0|H_0| 0 \rangle = A^2 E_s + (1-A^2) E_p
\label{e0}
\end{eqnarray}
where $H_0$ is the hydrogen atom Hamiltonian where $E_s \approx -19.38$ eV is the energy of the $2s$-state and $E_p \approx -11.07$ eV is the energy of the $2p$-state.

\begin{figure}[tbh]
\includegraphics[width=6cm, keepaspectratio]{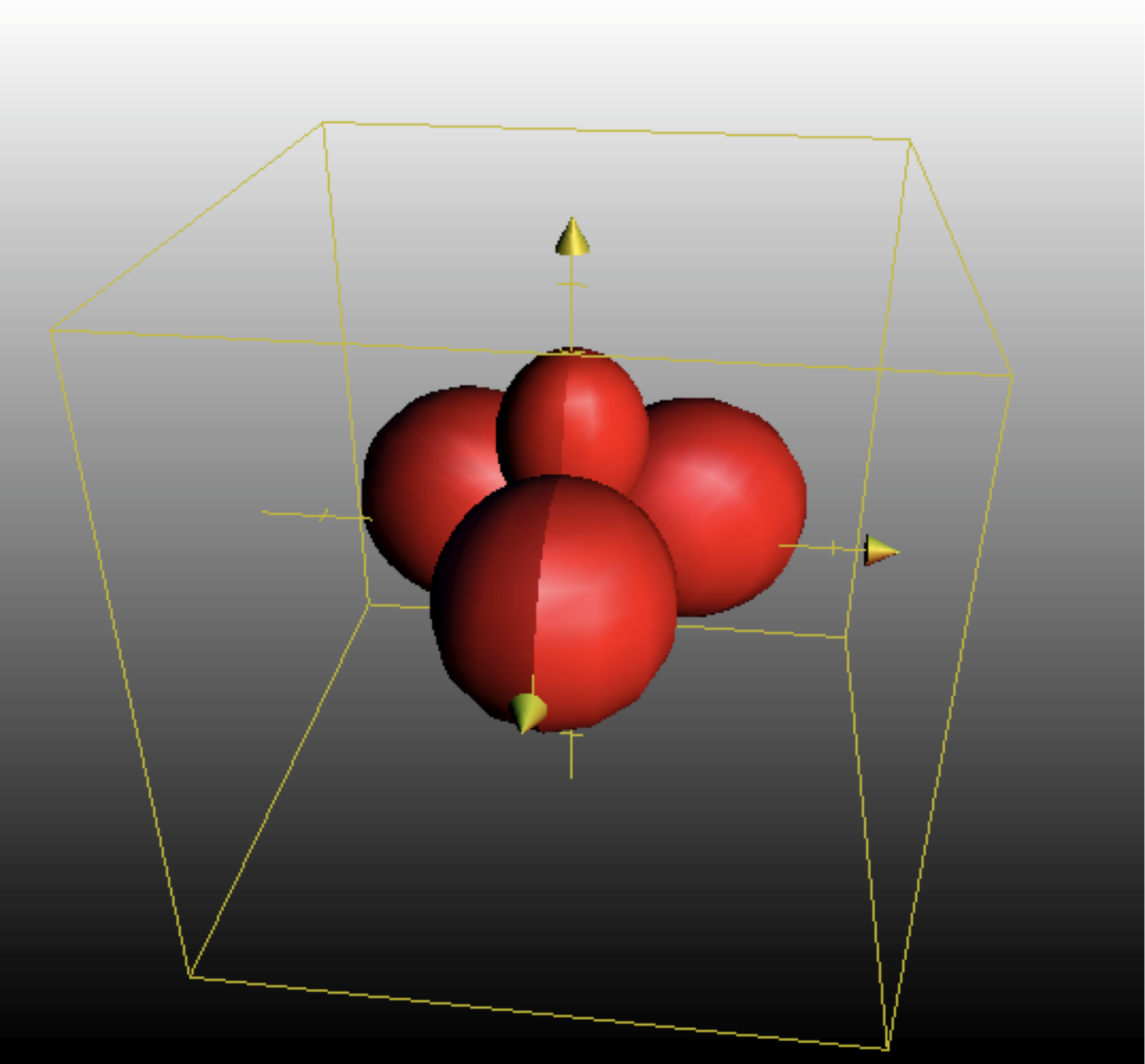}
\includegraphics[width=6cm, keepaspectratio]{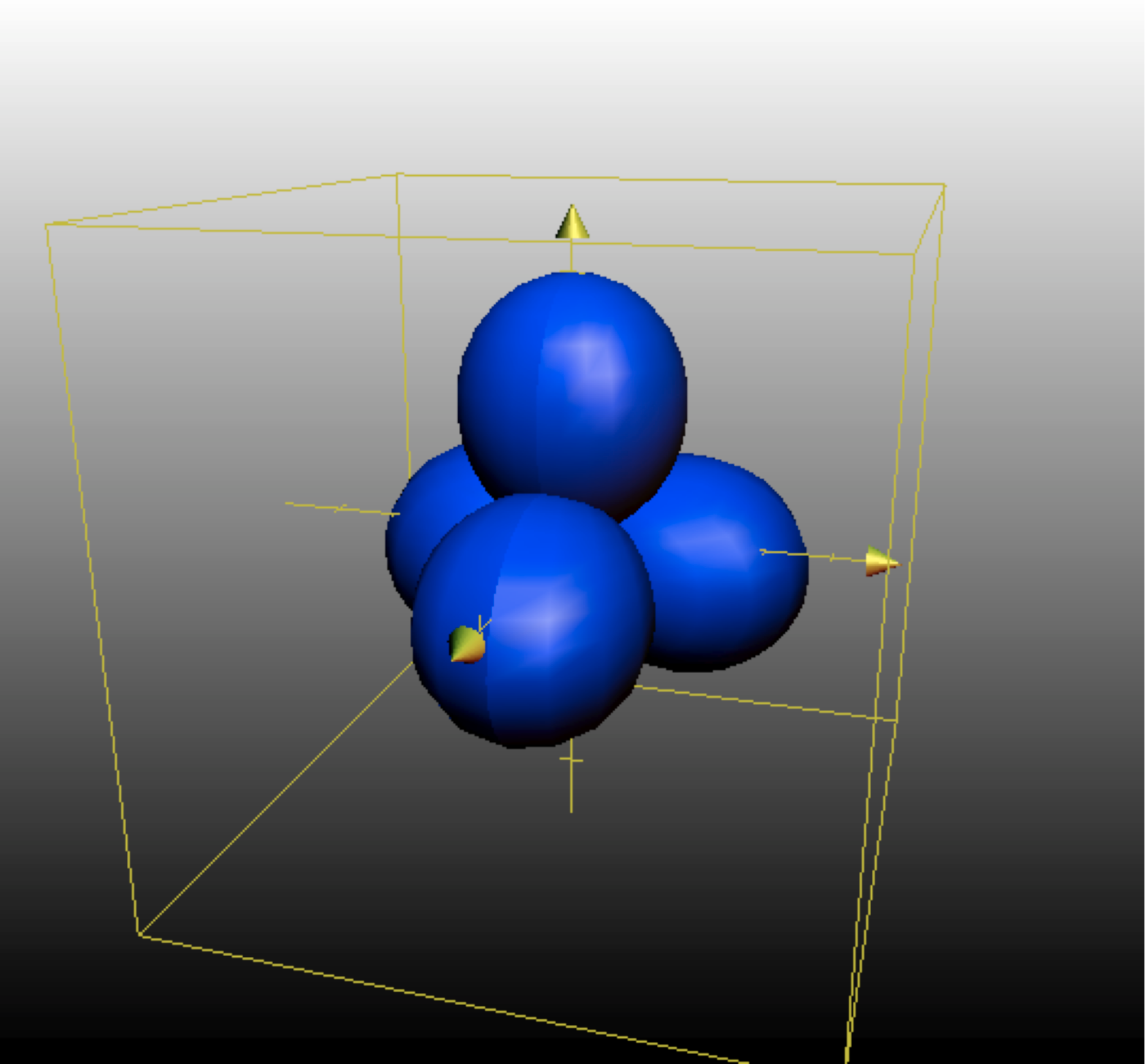}
\caption{$sp^2$ (left) and $sp^3$ (right) hybridized orbitals.}
\label{sp23fig}
\end{figure}

All the other 3 orthogonal orbitals can be constructed starting from (\ref{sp0}). A particularly simple parametrization is the following \cite{sographene}:
\begin{eqnarray} 
|1\rangle &=& \sqrt{(1-A^2)/3} |s\rangle + \sqrt{2/3} |p_x \rangle - (A/\sqrt{3})|p_z\rangle \, ,
\nonumber
\\
|2\rangle &=& \sqrt{(1-A^2)/3} |s\rangle - \sqrt{1/6} |p_x \rangle - \sqrt{1/2} |p_y\rangle - (A/\sqrt{3})|p_z\rangle \, , 
\nonumber
\\
|3\rangle &=& \sqrt{(1-A^2)/3} |s\rangle - \sqrt{1/6} |p_x \rangle + \sqrt{1/2} |p_y\rangle - (A/\sqrt{3})|p_z\rangle \, .
\label{hybrids}
\end{eqnarray}
Notice that $A$ controls the angle between the $z$ axis and these states. We can clearly see that the direction of largest amplitude for one of these orbitals (say, $\langle {\bf r} |1\rangle$) is given by:
\begin{eqnarray}
\frac{\partial}{\partial \theta} \langle {\bf r} | 1 \rangle (\phi=0) 
&=& \sqrt{2} \cos\theta_{\rm m} + A \sin\theta_{\rm m} = 0 \, ,
\nonumber
\\
\theta_{\rm m} &=& - \arctan(\sqrt{2}/A) \, . 
\end{eqnarray}
Hence, for $A=0$ the hybridized state $|1\rangle$ is perpendicular to the other orbitals that remain in the $x-y$ plane. This is the so-called $sp^2$ hybridization (see Fig.\ref{sp23fig}). For $A=1/2$ the orbitals have tetragonal structure making an angle of $109.47$ degrees with the $z$ axis. This is the $sp^3$ hybridization (see Fig.\ref{sp23fig}).

\begin{figure}[tbh]
\centerline{\includegraphics[width=8cm, keepaspectratio]{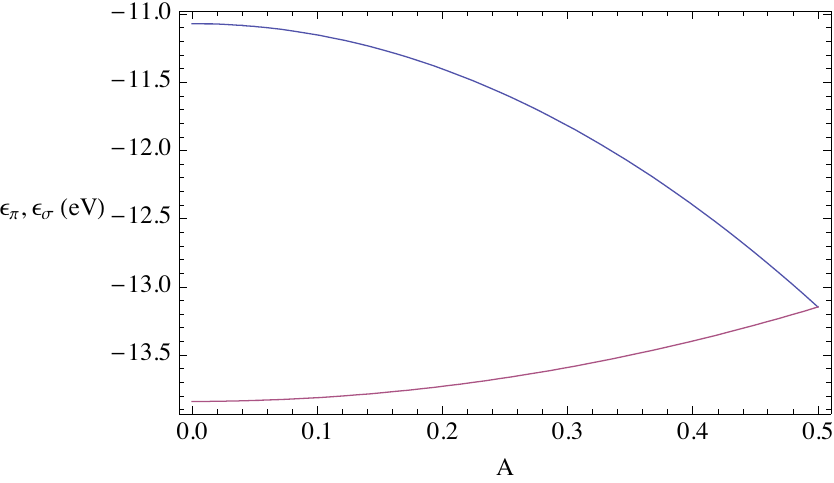}}
\caption{Energy of the hybridize $s-p$ states, given by (\ref{e0}) and (\ref{e1}) as a function of the hybridization parameter $A$. $A=0$ corresponds to the sp$^2$ and $A=1/2$ corresponds to the sp$^3$ configuration.}
\label{esep}
\end{figure}

In free space, the states $|1\rangle$, $|2 \rangle$ and $|3\rangle$ 
are clearly degenerate and their energy is given by:
\begin{eqnarray}
E_1=E_2=E_3= \langle 1|H_0|1 \rangle = \epsilon_{\sigma}= \frac{1-A^2}{3} E_s + \frac{2 + A^2}{3} E_p.
\label{e1}
\end{eqnarray}
The energy of these states is shown in Fig. \ref{esep}. Notice that
in the $sp^3$ case ($A=1/2$) all orbitals are degenerated while in
the $sp^2$ case the orbitals are separated by an energy of approximately
$2.77$ eV. 

\begin{figure}[tbh]
\centerline{\includegraphics[width=8cm, keepaspectratio]{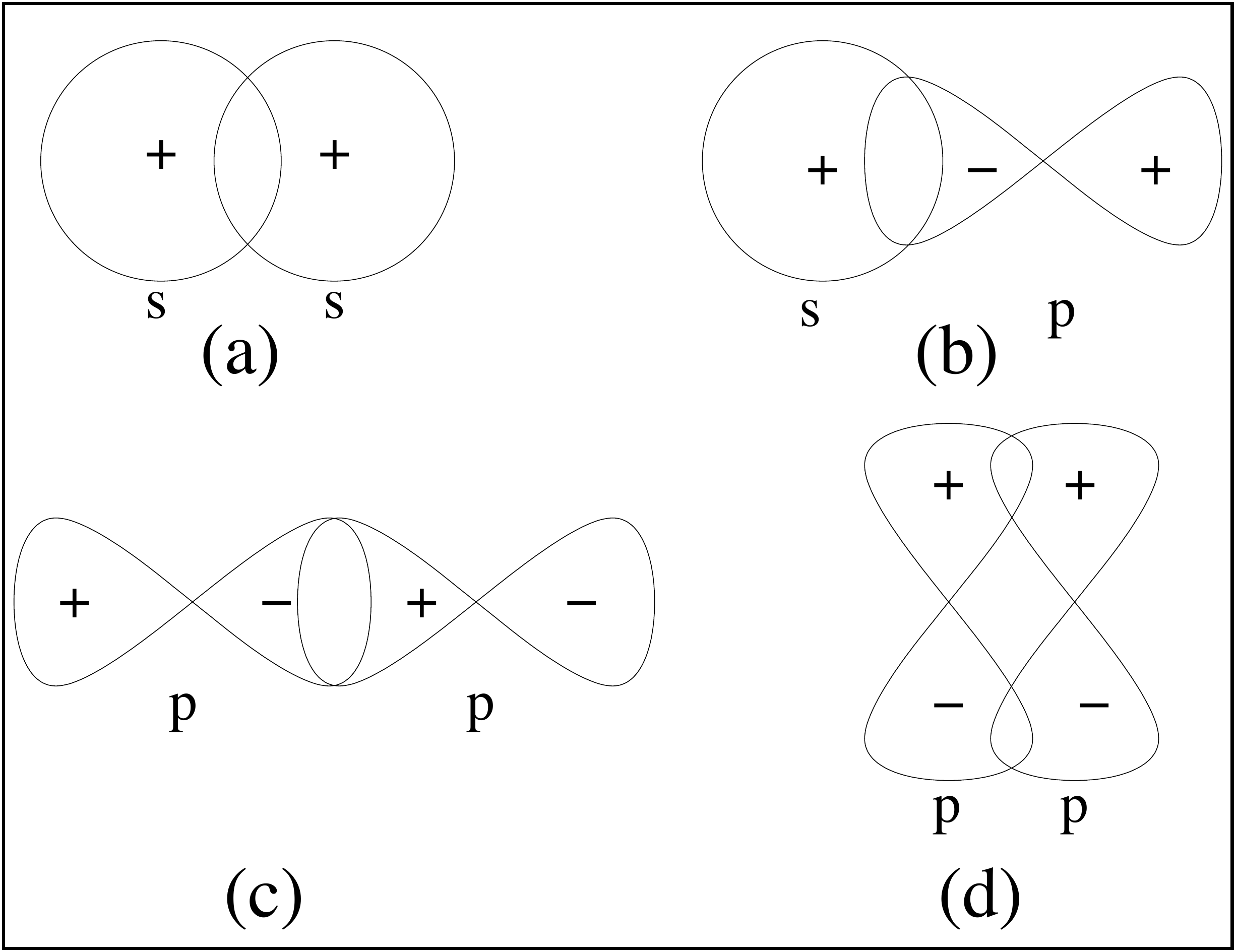}}
\caption{Basic hybridization energies for $s-p$ bonds: (a) $V_{ss\sigma}$; (b) $V_{sp\sigma}$; (c) $V_{pp\sigma}$; (d) $V_{pp\pi}$.}
\label{hybridizations}
\end{figure}

The presence of another carbon atom induces a hybridization between the different orbitals. This hybridization depends on the distance $\ell$ between the atoms and also on the orientation of the orbitals relative to each other. The distance dependence is usually well described by an exponential behavior:
\begin{eqnarray}
V_{\alpha}(\ell) \approx V^0_{\alpha} e^{-\kappa_{\alpha} \ell} \, ,
\end{eqnarray}
where $\kappa_{\alpha} = d \ln(V_{\alpha})/d\ell$ where $\alpha$ labels the different orientations of the orbitals.  In terms of orientation, there are four different types of elementary hybridization between different orbitals (shown in Fig.\ref{hybridizations}): $V_{ss\sigma}$ ($\approx -5$ eV for $\ell=1.42$ \, \AA); $V_{sp\sigma}$ ($\approx + 5.4$ eV for $\ell=1.42$ \, \AA); $V_{pp\sigma}$ ($\approx +8.4$ eV for $\ell=1.42$ \, \AA); $V_{pp\pi}$ ($\approx -2.4$ eV for $\ell=1.42$ \, \AA) \cite{harrison}.

\begin{figure}[tbh]
\centerline{\includegraphics[width=8cm,keepaspectratio]{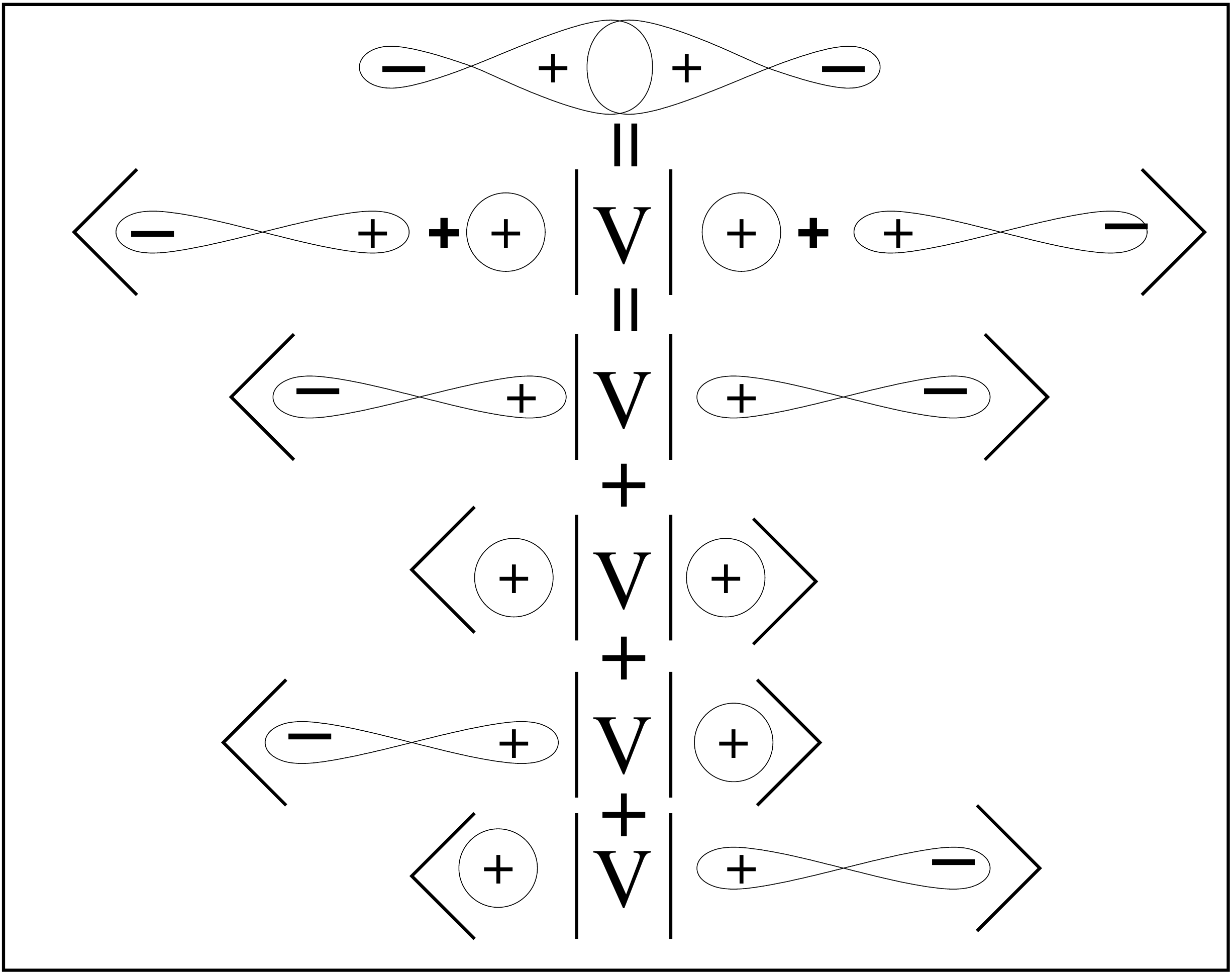}}
\caption{Calculation of the hybridization of two $s-p$ orbitals in terms of the basic hybridization energies shown in Fig.\ref{hybridizations}.}
\label{decomposition}
\end{figure}

Any hybridization energy can be obtained from those basic hybridizations shown in Fig. \ref{hybridizations} as linear combinations. For instance, let us consider the intra-atomic hybridization between between $|2 \rangle$ and $|3 \rangle$. This is given by the matrix element:
\begin{eqnarray}
V_{{\rm intra}} &=& \langle 2|H_0|3 \rangle = \langle 2| \left(\sqrt{\frac{1-A^2}{3}} E_s |s \rangle - \frac{E_p}{\sqrt{6}} |p_x \rangle + \frac{E_p}{\sqrt{2}} |p_y\rangle- \frac{A}{\sqrt{3}} E_p |p_z\rangle\right)
\nonumber
\\
&=& \frac{1-A^2}{3} (E_s - E_p) \, .
\label{v23}
\end{eqnarray}   
We can also compute inter-atomic hybridization energies such as the hybridization between two $|2\rangle$ states oriented as in Fig. \ref{decomposition}:
\begin{eqnarray}
V_{\sigma} = - \frac{2}{3} V_{pp\sigma} + \frac{1-A^2}{3} V_{ss\sigma} 
- \frac{2}{3} \sqrt{2(1-A^2)} V_{sp\sigma} \, .
\end{eqnarray}

\section{The Crystal and Band Structure}
\label{bands}

\begin{figure}[tbh]
\centerline{\includegraphics[width=10cm,keepaspectratio]{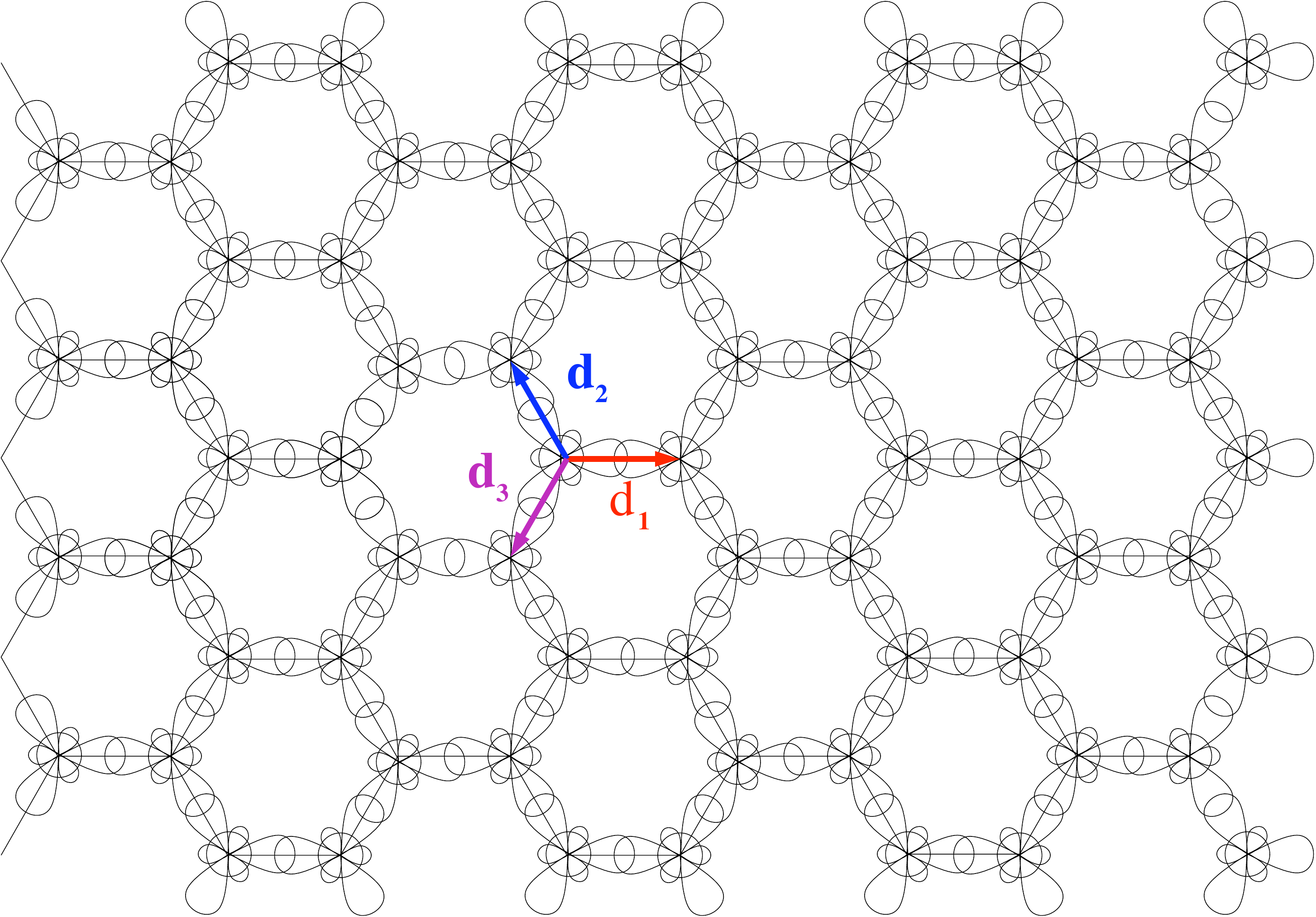}}
\caption{Honeycomb lattice and electronic orbitals}
\label{rede}
\end{figure}

\begin{figure}[tbh]
\centerline{\includegraphics[width=8cm, keepaspectratio]{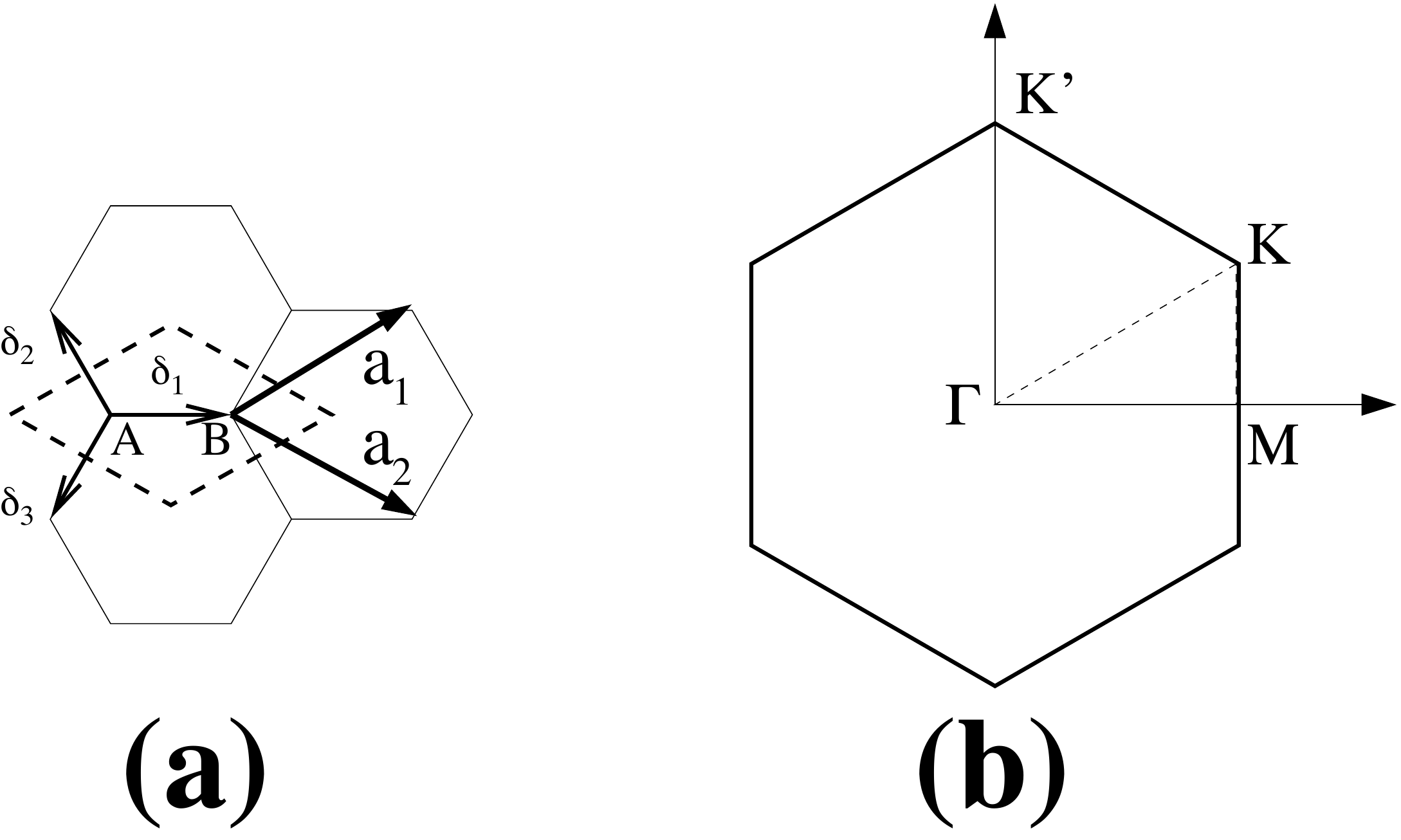}}
\caption{(a) Lattice structure; (b) First Brillouin zone}
\label{lattice}
\end{figure}

In this section we focus on the flat graphene ($sp^2$ only) lattice. The geometry of the lattice is shown in Fig.\ref{lattice} (a) and it can be easily seen that the honeycomb lattice
can be considered a crystal lattice with two atoms per unit cell. These two sublattices, $A$ and $B$, form two interpenetrating triangular lattices. The lattice vectors are:
\begin{eqnarray}
{\bf a}_1 &=& \frac{a}{2} (3, \sqrt{3}) \, ,
\nonumber
\\
{\bf a}_2 &=& \frac{a}{2} (3, -\sqrt{3}) \, ,
\end{eqnarray}
and the nearest neigh boor vectors are:
\begin{eqnarray}
\vec{\delta}_1 &=& a (1,0) \, ,
\nonumber
\\
\vec{\delta}_2 &=& \frac{a}{2} (-1,\sqrt{3}) \, ,
\nonumber
\\
\vec{\delta}_3 &=& \frac{a}{2} (-1,-\sqrt{3}) \, ,
\label{deltat}
\end{eqnarray}
where $a = 1.42$ \AA \, is the carbon-carbon distance. The first Brillouin zone is shown in Fig. \ref{lattice}(b) and the reciprocal lattice is spanned by the vectors:
\begin{eqnarray}
{\bf b}_1 &=& \frac{2 \pi}{3 a} (1, \sqrt{3}) \, ,
\nonumber
\\
{\bf b}_2 &=& \frac{2 \pi}{3} (1, -\sqrt{3}) \, .
\label{recvec}
\end{eqnarray}
The $K$ point, located at 
\begin{eqnarray}
{\bf K} = \frac{2 \pi}{3 a} \left(1,\frac{1}{\sqrt{3}}\right)
\label{kpoint}
\end{eqnarray}
is of particular importance here. 

Let us label the state at each carbon atom as: $|n,a,i\rangle$ where $n=1,...,N$ labels the unit cell, $a=A,B$ labels the sub-lattice (see Fig.\ref{lattice}(a)), and $i=0,1,2,3$ labels the states defined in (\ref{sp0}) and (\ref{hybrids}). 
In the simplest approximation we can consider a tight binding description where the electrons hop between different orbitals (intraband) in the same atom with characteristic energy $V_{{\rm intra}}$, between the two different
sublattices (interband) along the planar orbitals with energy $V_{\sigma}$, and between two different sublattices with orbitals perpendicular to the plane with energy $V_{pp\pi}$. In this case the Hamiltonian can be written as:   
\begin{eqnarray}
H &=& \epsilon_{\pi} \sum_{n,a} c^{\dag}_{n,a,0}c_{n,a,0} + \epsilon_{\sigma} \sum_{n,a,i \neq 0} c^{\dag}_{n,a,i}c_{n,a,i} + V_{{\rm intra}} \sum_{n,a,i \neq j \neq 0} 
(c^{\dag}_{n,a,i} c_{n,a,j} + {\rm h.c.}) 
\nonumber
\\
&+& V_{pp\pi} \sum_{\langle n,m \rangle} (c^{\dag}_{n,A,0} c_{m,B,0} + {\rm h.c.}) + V_{\sigma} \sum_{\langle n,m \rangle,i \neq 0} (c^{\dag}_{n,A,i} c_{m,B,i} + {\rm h.c.}) \, .
\label{tbhamil}
\end{eqnarray}
The first step for the solution of this problem is to use the fact that the system is invariant under discrete translations and to Fourier transform the operators:
\begin{eqnarray}
c_{n,a,i} = \frac{1}{\sqrt{N}} \sum_{{\bf k}} e^{i {\bf k} \cdot {\bf R}_n} c_{{\bf k},a,i} \, ,
\end{eqnarray}
which leads to:
\begin{eqnarray}
H &=& \sum_{{\bf k}} \left\{ \epsilon_{{\pi}} \sum_a c^{\dag}_{{\bf k},a,0} c_{{\bf k},a, 0} + \epsilon_{\sigma} \sum_{a,i \neq 0} c^{\dag}_{{\bf k},a,i} c_{{\bf k},a,i} + V_{{\rm intra}} \sum_{a,i \neq j \neq 0} (c^{\dag}_{{\bf k},a,i} c_{{\bf k},a,j} + {\rm h.c.})  \right.
\nonumber
\\
&+& \left. V_{pp\pi} \gamma_{{\bf k}} (c^{\dag}_{{\bf k},A,0} c_{{\bf k},B,0} + {\rm h.c.}) 
+ V_{\sigma} \left[e^{i {\bf k} \cdot \vec{\delta}_1} c^{\dag}_{{\bf k},A,1} c_{{\bf k},B,1}+ e^{i {\bf k} \cdot \vec{\delta}_2} c^{\dag}_{{\bf k},A,2} c_{{\bf k},B,2} + e^{i {\bf k} \cdot \vec{\delta}_3} c^{\dag}_{{\bf k},A,3} c_{{\bf k},B,3} + {\rm h.c.} \right] \right\}
\label{basichamil}
\end{eqnarray}
where
\begin{eqnarray}
\gamma_{{\bf k}} &=& \sum_{i=1,2,3} e^{i {\bf k} \cdot \vec{\delta}_i} \, ,
\nonumber
\\
|\gamma_{{\bf k}}| &=& \sqrt{3 + 2 \cos({\bf k} \cdot (\vec{\delta}_1-\vec{\delta}_2))
+ 2 \cos({\bf k} \cdot (\vec{\delta}_1-\vec{\delta}_3)) + 2 \cos({\bf k} \cdot (\vec{\delta}_2-\vec{\delta}_3))} \, .
\end{eqnarray}
In this case the Hamiltonian can be written as:
\begin{eqnarray}
H = \sum_{{\bf k}} \Psi^{\dag}_{{\bf k}} \cdot [H] \cdot \Psi_{{\bf k}} \, ,
\end{eqnarray}
where $\Psi^{\dag}_{{\bf k}}=(c^{\dag}_{{\bf k},A,0},c^{\dag}_{{\bf k},B,0},c^{\dag}_{{\bf k},A,1},c^{\dag}_{{\bf k},B,1},c^{\dag}_{{\bf k},A,2},c^{\dag}_{{\bf k},B,2},c^{\dag}_{{\bf k},A,3},c^{\dag}_{{\bf k},B,3})$ and
\begin{eqnarray}
[H] = \left[\begin{array}{cccccccc}
\epsilon_{\pi} & V_{pp\pi} \gamma_{{\bf k}} & 0 & 0 & 0 & 0 & 0 & 0 \\
V_{pp\pi} \gamma^*_{{\bf k}} & \epsilon_{\pi} & 0 & 0 & 0 &0 & 0 & 0 \\
0 & 0 & \epsilon_{\sigma} & V_{\sigma} e^{i {\bf k} \cdot \vec{\delta}_1} & V_{{\rm intra}} & 0 & V_{{\rm intra}} & 0 \\
0 & 0 & V_{\sigma} e^{- i {\bf k} \cdot \vec{\delta}_1} &\epsilon_{\sigma} &  0 & V_{{\rm intra}} & 0 & V_{{\rm intra}} \\
0 & 0 & V_{{\rm intra}} & 0 & \epsilon_{\sigma} & V_{\sigma} e^{i {\bf k} \cdot \vec{\delta}_2} & V_{{\rm intra}} & 0 \\
0 & 0 & 0 & V_{{\rm intra}} & V_{\sigma} e^{-i {\bf k} \cdot \vec{\delta}_2} & \epsilon_{\sigma} & 0 & V_{{\rm intra}} \\
0 & 0 & V_{{\rm intra}} & 0 & V_{{\rm intra}} & 0 & \epsilon_{\sigma} & V_{\sigma} e^{i {\bf k} \cdot \vec{\delta}_3} \\
0 & 0 & 0 & V_{{\rm intra}} & 0 & V_{{\rm intra}} & V_{\sigma} e^{-i {\bf k} \cdot \vec{\delta}_3} & \epsilon_{\sigma} \\
\end{array} \right] \, .
\label{hamil}
\end{eqnarray}
It is immediately obvious that the Hamiltonian for the $\pi$ band completely decouples from the $\sigma$ band and can
be treated separately.

\begin{figure}[tbh]
(a) \centerline{\includegraphics[width=8cm,keepaspectratio]{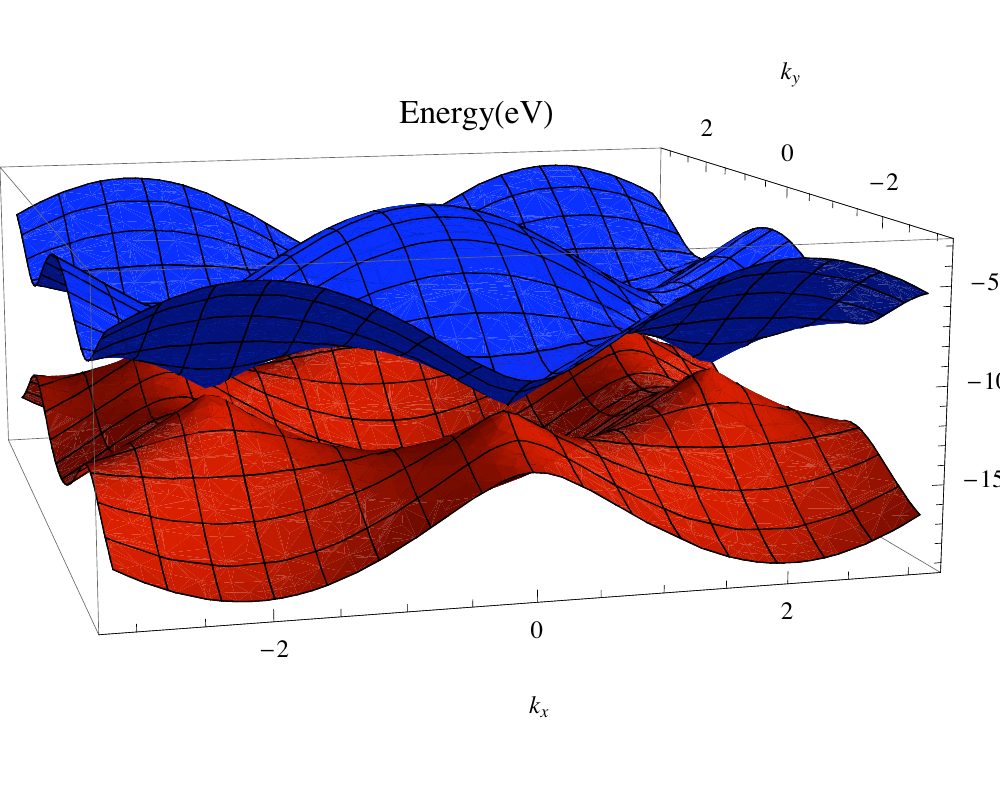}}
(b) \centerline{\includegraphics[width=8cm,keepaspectratio]{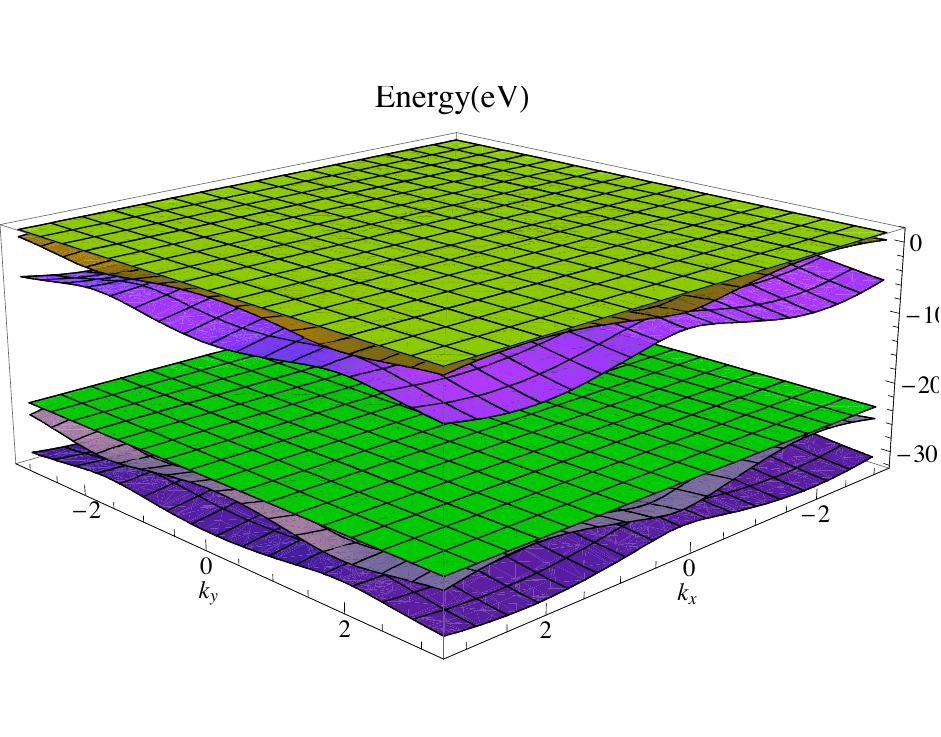}}
\caption{Band structure of graphene (energy in ev). (a) $\pi$-bands; (b) $\sigma$-bands.}
\label{pisig}
\end{figure}

\begin{figure}[tbh]
\centerline{\includegraphics[width=16cm,keepaspectratio]{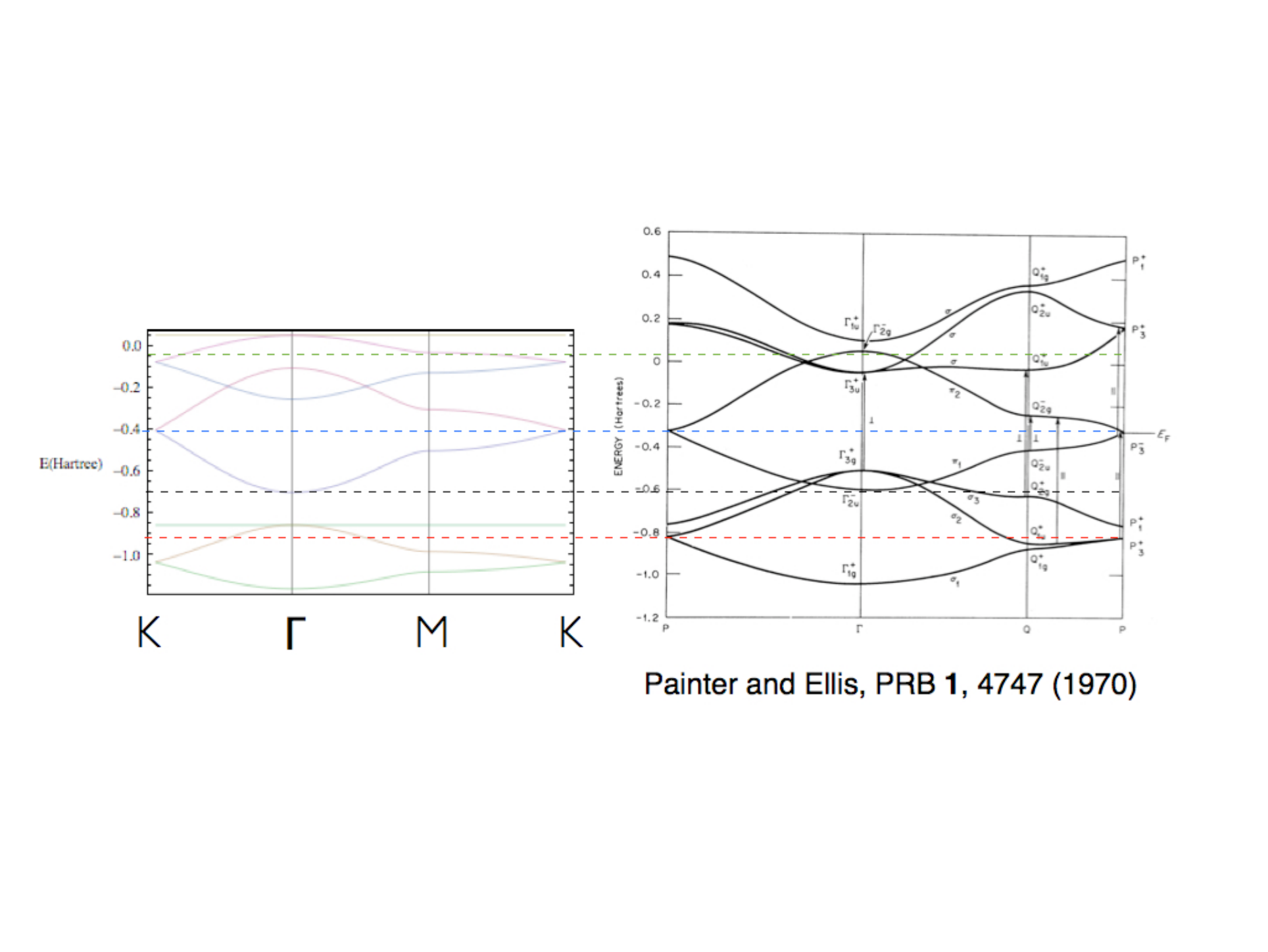}}
\caption{Band structure of graphene (energy is given in Hartrees - 1 Hartree = 27.21 eV): left, result of the model discussed in the text; right, result of ref. [\onlinecite{painter}].}
\label{bandstruct}
\end{figure}

Although (\ref{hamil}) is a $8 \times 8$ matrix (a $2 \times 2$ block plus a $6 \times 6$ block), it can be diagonalized analytically to produce $8$ energy bands that read:
\begin{eqnarray}
E_{\pi,\pm}({\bf k}) &=& \epsilon_{\pi} \pm |V_{pp\pi}| |\gamma_{{\bf k}}| \, ,
\nonumber
\\
E_{\sigma,1,\pm}({\bf k}) &=& \epsilon_{\sigma} - V_{{\rm intra}} \pm V_{\sigma} \, ,
\nonumber
\\
E_{\sigma,2,\pm}({\bf k}) &=& \epsilon_{\sigma} + \frac{V_{{\rm intra}}}{2} + \sqrt{\left(\frac{3 V_{{\rm intra}}}{2}\right)^2 + V_{\sigma}^2 \pm |V_{{\rm intra}} V_{\sigma}| |\gamma_{{\bf k}}}| \, ,
\nonumber
\\
E_{\sigma,3,\pm}({\bf k}) &=& \epsilon_{\sigma} + \frac{V_{{\rm intra}}}{2} - \sqrt{\left(\frac{3 V_{{\rm intra}}}{2}\right)^2 + V_{\sigma}^2 \pm |V_{{\rm intra}} V_{\sigma}| |\gamma_{{\bf k}}}| \, ,
\label{exa}
\end{eqnarray}
which represent two $\pi$ bands and six $\sigma$ bands, respectively. 
The $\pi$-bands are shown in Fig.\ref{pisig}(a) and the $\sigma$-bands are shown in Fig. \ref{pisig}(b). 
In Fig.\ref{bandstruct} we compare the results of this simple tight-binding with more sophisticated calculations \cite{painter}. One can clearly see that the tight binding approach produces
a fairly good description of the band structure, especially the $\pi$ bands, although more hopping parameters have to be introduced in order to reproduce the details 
of the $\sigma$ bands. The fact that this full band structure can be obtained analytically makes this particular parametrization of the bands rather attractive as a first step towards
the electronic description of the graphene electrons. 

As we said in Section \ref{chemistry}, the carbon $1 s$ states are completely filled, leaving 4 electrons per carbon atom (8 per unit cell) to fill these 8 bands. Hence, the first 4 bands are fully occupied meaning that the Fermi energy crosses exactly at the ``conical'' points of the dispersion at the K and K' point of the Brillouin zone in Fig.\ref{bandstruct}. Notice that there are 6 of these points in the Brillouin zone with vectors given by:
\begin{eqnarray}
{\bf Q}_1 &=& 4 \pi/(3 \sqrt{3} a) (0,1) \, ,
\label{vecQ}
\\
{\bf Q}_2 &=& 4 \pi/(3 \sqrt{3} a) (\sqrt{3}/2,1/2) \, ,
\nonumber
\\
{\bf Q}_3 &=& 4 \pi/(3 \sqrt{3} a) (\sqrt{3}/2,-1/2) \, ,
\nonumber
\end{eqnarray}
and also at $-{\bf Q}_i$ with $i=1,2,3$. Close to these particular points, that is, close to the Fermi energy, 
we can find a simple expression for the electronic dispersion by expanding $\gamma_{{\bf k}}$ as:
\begin{eqnarray}
\gamma_{{\bf Q}_1 + {\bf k}} \approx \frac{3 a}{2} (k_x + i k_y) \, ,
\end{eqnarray}
for $k \ll Q$. From (\ref{exa}) we see that the spectrum becomes:
\begin{eqnarray}
E_{\pm}(k_x,k_y) = \pm v_F k = \pm v_F \sqrt{k_x^2+k_y^2}\, ,
\end{eqnarray}
where 
\begin{eqnarray}
v_F = 3 |V_{pp\pi}| a/2 \, ,
\end{eqnarray}
is the Fermi velocity. Thus the electronic spectrum close to the Fermi energy has a conical form, 
mimicking the dispersion of a relativistic, massless, Dirac particle. 
Notice that in the first Brillouin zone each corner of the zone contains $1/3$ of a Dirac cone but 
we do not need all the vectors in (\ref{vecQ}) to describe the problem. We can use reciprocal lattice
vectors (\ref{recvec})  in order to translate each piece of the cone to two corners
located at $\pm Q_1$ making 2 Dirac cones in the extended zone scheme.
Hence, close to ${\bf Q}_1$ we can rewrite the Hamiltonian as:
\begin{eqnarray}
H_0 \approx v_F \sum_{{\bf k}} 
(\psi^{\dag}_{A,{\bf k}},\psi^{\dag}_{B,{\bf k}} )
\cdot 
\left[
\begin{array}{cc}
0 \hspace{0.5cm} & k_x + i k_y \\
k_x - i k_y \hspace{0.5cm} & 0
\end{array}
\right]
\cdot
\left(
\begin{array}{c}
\psi_{A,{\bf k}} \\
\psi_{B,{\bf k}}
\end{array}
\right) \, ,
\end{eqnarray}
where $\psi_{A,{\bf k}} = a_{{\bf Q}_1 +{\bf k}}$ and similarly for the
$B$ sublattice. By Fourier transforming the Hamiltonian back to real space we obtain the 2D Dirac Hamiltonian:
\begin{eqnarray}
H_0 = \int d^2 r \Psi^{\dag}({\bf r}) (i v_F \vec{\sigma} \cdot \nabla
-\mu) \Psi({\bf r}) \, ,
\label{dirac}
\end{eqnarray}
where $\mu$ is the chemical potential measured away from the Dirac point, $\vec{\sigma}= (\sigma_x,\sigma_y)$
are Pauli matrices, and
\begin{eqnarray}
\Psi^{\dag}({\bf r}) = (\psi^{\dag}_{A}({\bf r}),\psi^{\dag}_{B}({\bf r})) \, .
\end{eqnarray}

\section{Phonons in Free Floating Graphene}
\label{phonons}

\begin{figure}[tbh]
\centerline{\includegraphics[width=8cm,keepaspectratio]{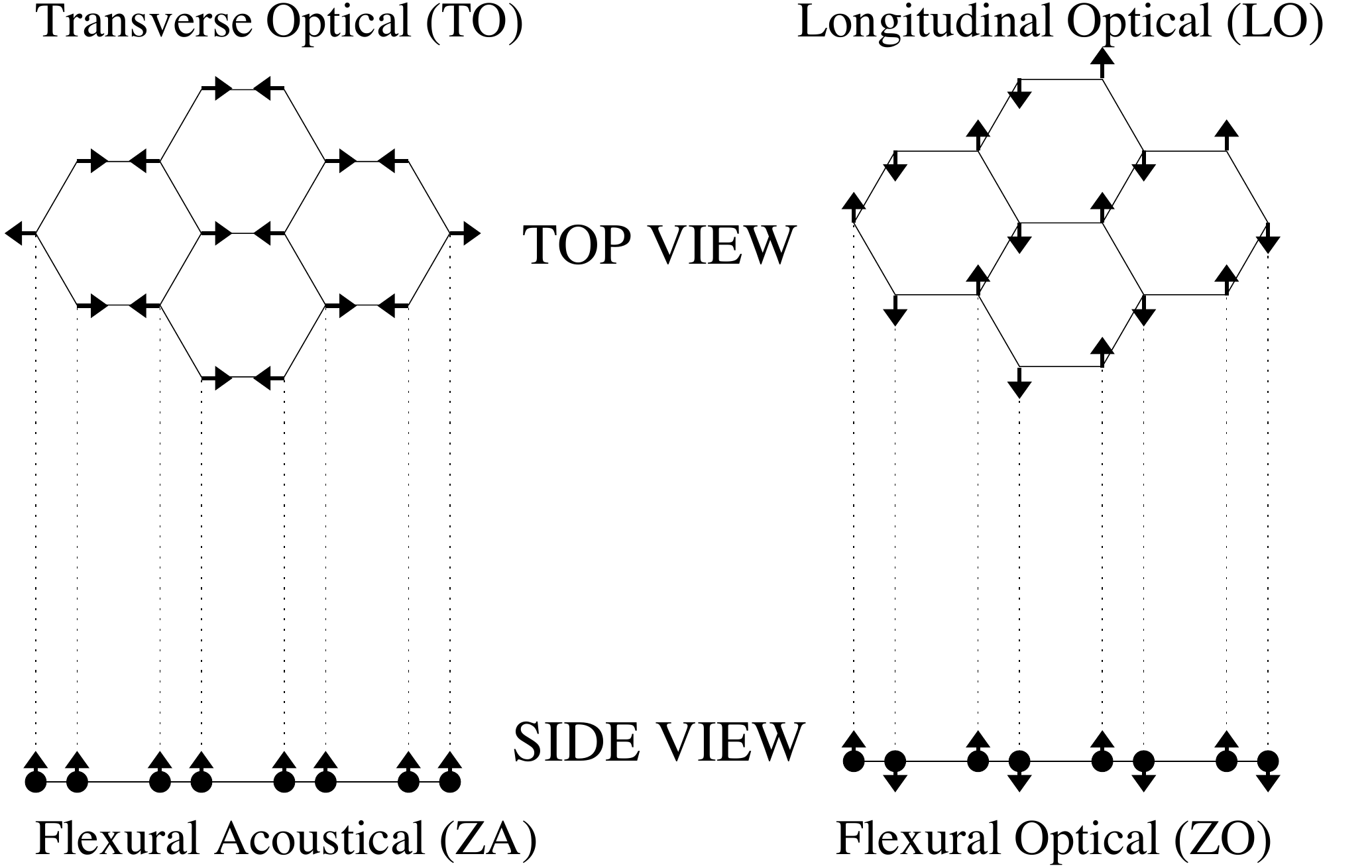}}
\caption{Lattice displacements associated with some of the long-wavelength
phonon modes in graphene.}
\label{phononsdisp}
\end{figure}

Since graphene is two-dimensional and has 2 atoms per unit cell there are 4 in-plane degrees of freedom that give rise to 2 acoustic,
longitudinal (LA) and transverse (TA), and 2 optical, longitudinal (LO) and transverse (TO), phonon modes. However, graphene is embedded in
a three-dimensional space and hence there are two extra degrees of freedom associated with out-of-plane motion: flexural acoustical (ZA) 
and optical (ZO) phonon modes. The displacements associated with  these modes at long wavelengths is shown in Fig.\ref{phononsdisp}. 
In leading order in the displacements the in-plane and out-of-plane modes decouple and we can study them separately. 
Here we focus in the out-of-plane modes since those dominate the low energy physics of graphene. 

We use the Monge representation where the a point ${\bf R}$ in the graphene lattice is described by the vector ${\bf R} = ({\bf r},h({\bf r}))$
where ${\bf r}=(x,y)$ is the 2D coordinate vector and $h({\bf r})$ is the height variable \label{rpm09}. The unit vector normal
to the surface is given by:
\begin{eqnarray}
{\bf N} = (-\nabla h + {\bf z})/\sqrt{1+(\nabla h)^2},
\label{normal}
\end{eqnarray}
where $\nabla = (\partial_x,\partial_y)$ is the 2D gradient operator, and ${\bf z}$ is the unit vector in the third direction. Distortions from the flat configuration cost energy because they rotate the sigma orbitals that want to be aligned in order to get maximum overlap of the wavefunctions that bind the atoms together. Let us consider the triangular lattice centered at sublattice A, say at ${\bf R}_i$, that covers the entire graphene lattice. The normal at this point is ${\bf N}_i$. The energy cost for the bending of graphene can then be written as:
\begin{eqnarray}
U_B = - \frac{\kappa_L}{2} \sum_{\langle i,j \rangle} {\bf N}_i \cdot {\bf N}_j \, ,
\label{bend}
\end{eqnarray}
where $\kappa_L$ is the lattice bending rigidity of graphene. Notice that (\ref{bend}) has the form of the energy of a classical ferromagnet 
where the normal vectors play the role of spins. Define ${\bf R}_j = {\bf R}_i + {\bf u}$ where ${\bf u}$ is the next-to-nearest 
neighbor vector that connects two triangles and let us consider only length scales much larger than the lattice spacing, 
$|{\bf u}| \approx \sqrt{3} a$. In this, assuming that ${\bf N}$ is a smooth function over the lattice scale, we can write:
\begin{eqnarray}
{\bf N}_j \approx {\bf N}_i + ({\bf u} \cdot \nabla) {\bf N}_i +
\frac{1}{2} ({\bf u} \cdot \nabla)^2 {\bf N}_i
\end{eqnarray}
and hence:
\begin{eqnarray}
{\bf N}_i \cdot {\bf N}_j \approx 1 + {\bf N}_i \cdot [({\bf u} \cdot \nabla) {\bf N}_i] +
\frac{1}{2} {\bf N}_i \cdot [({\bf u} \cdot \nabla)^2 {\bf N}_i]
\label{mainexp}
\end{eqnarray}
where we used ${\bf N}_i^2 =1$. Again, assuming the distortions to be
small and smooth, we can rewrite (\ref{normal}) as:
\begin{eqnarray}
{\bf N} \approx {\bf z} - \nabla h - \frac{{\bf z}}{2} (\nabla h)^2 \, ,
\label{ncont}
\end{eqnarray}
and hence, 
\begin{eqnarray}
\partial_a {\bf N} \approx - \partial_a \nabla h - {\bf z} \nabla h \cdot
(\partial_a \nabla h) \, ,
\label{dncont}
\end{eqnarray}
where $a = x,y$. Therefore, using (\ref{ncont}) and (\ref{dncont}) and the
fact that ${\bf z} \cdot \nabla h = 0$, we find:
\begin{eqnarray}
{\bf N} \cdot [\partial_a {\bf N}] \approx 0 \, .
\label{nlin}
\end{eqnarray}
Once again:
\begin{eqnarray}
\partial^2_{a,b} {\bf N} \approx 
- \partial^2_{a,b} \nabla h - {\bf z} (\partial_n
\nabla h) \cdot (\partial_a \nabla h) - {\bf z} \nabla h \cdot
(\partial^2_{a,b} \nabla h) \, ,
\end{eqnarray}
and finally:
\begin{eqnarray}
{\bf N} \cdot [\partial^2_{a,b} {\bf N}] 
\approx - (\partial_b \nabla h) \cdot (\partial_a \nabla h) \, .
\label{nsq}
\end{eqnarray}
Replacing (\ref{nlin}) and (\ref{nsq}) into (\ref{mainexp}) we find:
\begin{eqnarray}
{\bf N}_i \cdot {\bf N}_j \approx 1 - \frac{1}{2} \left[({\bf u} \cdot
  \nabla) \nabla h\right]^2 \, .
\label{over}
\end{eqnarray}
Replacing (\ref{over}) in (\ref{bend}) we find:
\begin{eqnarray}
\delta U_B &\approx& \frac{\kappa_L}{4} \sum_{i,{\bf u}} 
\left[({\bf u} \cdot \nabla) \nabla h({\bf R_i})\right]^2
\nonumber
\\
&\approx& \frac{\kappa}{2} \int d^2r \left[\nabla^2 h({\bf r})\right]^2 \, ,
\label{bendene}
\end{eqnarray}
where $\kappa$ ($\propto \kappa_L$) is the bending rigidity. Notice that
the energy (\ref{bendene}) is invariant under translations along the $z$
direction and rotations around any axis (the energy is invariant under 
$h \to h + C_1 a + C_2 {\bf r} \times \vec{\theta}$
where $C_1$ and $C_2$ are contains and $\vec{\theta}$ is the anti-clockwise oriented
angle of rotation along a given axis). In momentum space we can rewrite (\ref{bendene}) as:
\begin{eqnarray}
\delta U_B = \frac{\kappa}{2} \int d^2 q q^4 |h({\bf q})|^2 \, ,
\end{eqnarray}
showing the elastic cost in energy for bending the graphene sheet behaves as $q^4$ when $q \to 0$.

The quantum mechanics of bending can be easily obtained by quantization of the field $h({\bf r})$. 
Introducing a momentum operator $P({\bf q})$ that is canonically conjugated to $h({\bf q})$:
\begin{eqnarray}
\left[h({\bf q}),P(\bf q')\right]= i \delta^2({\bf q}-{\bf q'}) \, ,
\end{eqnarray}
we can write the Hamiltonian for the bending modes as:
\begin{eqnarray}
H = \int d^2 q \left\{ \frac{P(-{\bf q}) P({\bf q})}{2 \sigma} + \frac{\kappa q^4}{2}
h(-{\bf q}) h({\bf q})\right\} \, ,
\label{hph}
\end{eqnarray} 
where $\sigma$ is graphene's 2D mass density. From the Heisenberg equations of
motion for the operators it is trivial to find that $h({\bf q})$ oscillates harmonically
with a frequency given by:
\begin{eqnarray}
\omega_{{\rm flex}}({\bf q}) = \left(\frac{\kappa}{\sigma}\right)^{1/2} q^2 \, ,
\label{omegaflex}
\end{eqnarray}
which is the dispersion of the acoustical flexural mode.

Notice that the existence of a mode dispersing like $q^2$ in two dimensions has some
strong consequences. Consider, for instance, the mean square displacement of the height ($\omega_n = 2\pi n/\beta$ with $\beta = 1/T$ is the Matsubara frequency at temperature $T$):  
\begin{eqnarray}
\langle h^2 \rangle &=& \frac{1}{\beta} \sum_n \int \frac{d^2 q}{(2 \pi)^2} \frac{1}{(\omega_n)^2 +
  \kappa q^4}  
\nonumber
\\
&=& \int \frac{d^2 q}{(2 \pi)^2} \frac{coth\left(\frac{\beta \sqrt{\kappa} q^2}{2 \pi}\right)}{2 \sqrt{\kappa} q^2}
\nonumber
\\
&=& \frac{1}{8 \pi \sqrt{\kappa}} \int_{\beta \sqrt{\kappa}/(2 \pi L^2)}^{\beta \sqrt{\kappa} \Lambda^2/(2 \pi)}
du \frac{coth(u)}{u}
\end{eqnarray}
where we had to introduce a ultraviolet cut-off ($\Lambda \sim 1/a$) and a infrared cut-off ($1/L$ where $L$
is the system size) because the integral is formally divergent in both limits. The ultraviolet divergence
is not of concern since there is a physical cut-off which is the lattice spacing, the problem is the
infrared divergence since it indicates an instability in the system. Let us consider two different limits.
At low temperatures ($T \to 0$) we can approximate $coth(u) \approx 1$ and we get:
\begin{eqnarray}
\langle h^2 (T \to 0) \rangle \approx \frac{1}{4 \pi \sqrt{\kappa}} \ln(\Lambda L)
\end{eqnarray}
indicating that even at zero temperature the quantum fluctuations are logarithmically divergent. However,
at high $T$ we can approximate $coth(u) \approx 1/u$, the ultraviolet cut-off becomes irrelevant, and we find
\begin{eqnarray}
\langle h^2 (T \to \infty) \rangle \approx \frac{L^2}{4 \kappa \beta}  
\end{eqnarray} 
and one finds a severe infrared divergence of the thermal fluctuations. Similar arguments can be made
about the fluctuations of the normal ${\bf N}$, namely,
\begin{eqnarray}
\langle (\delta {\bf N})^2 \rangle \approx \langle (\nabla h)^2 \rangle \approx
\frac{1}{4 \kappa \beta} 
\ln\left[\frac{\sinh(\beta \sqrt{\kappa} \Lambda^2/(2 \pi))}{\sinh(\beta \sqrt{\kappa}/(2 \pi L^2))}\right] \, ,
\end{eqnarray}
and hence at low temperatures we find 
$\langle (\delta {\bf N})^2 (T \to 0) \rangle \approx \Lambda^2/(16 \pi \sqrt{\kappa})$ and the problem 
is free of infrared divergences. At high temperatures we have 
$\langle (\delta {\bf N})^2 (T \to \infty) \rangle \approx \ln[\Lambda L]/(2 \beta \kappa)$ and the fluctuations of the normal are logarithmically divergent. 
The interpretation of these results are straightforward: at zero temperature the graphene sheet should be rough but mostly flat, 
as one increases the temperature the thermal fluctuations become large and the graphene crumples (divergence of the normal indicates the formation of 
folds and creases). These results are modified by non-linear effects that we do not consider here \cite{chaikin}. More important than those, is the fact that
in most graphene experiments, where the measurement of electric properties depends of physically constraining the samples, the basic symmetries of free floating
graphene are explicitly broken. This is what we consider in the next Section. 

\section{Constrained Graphene}
\label{constraint}

Since graphene is not floating in free space and since there can be impurities that hybridize with
the carbon atoms changing the structural properties, it is important to consider symmetry breaking
process that change the picture presented in the last Section. Consider, for instance, if graphene is under tension.
In this case, the rotational symmetry along the $x$ and $y$ axis is broken and a new term is allowed
in the energy, that reads:
\begin{eqnarray}
U_T = \frac{\gamma}{2} \int d^2 r \, \left(\nabla h({\bf r})\right)^2 \, ,
\label{tension}
\end{eqnarray}
where $\gamma$ plays the role of the surface tension. It is easy to see that the dispersion is modified to:
\begin{eqnarray}
\omega({\bf q}) = q \, \sqrt{\frac{\kappa}{\sigma} q^2 + \frac{\gamma}{\sigma}}  \, ,
\label{omegat}
\end{eqnarray}
indicating that the dispersion of the flexural modes becomes linear in $q$, 
as $q \to 0$, under tension. Notice that this is the case of a 3D solid, that is, the flexural
mode becomes an ordinary acoustic phonon mode. 
This is what happens in graphite where the interaction between layers breaks the rotational
symmetry of the graphene layers. It is easy to see that in this case the graphene fluctuations
become suppressed: at low temperatures we find $\langle h^2 (T \to 0) \rangle$ is constant 
while $\langle h^2 (T \to \infty) \rangle \approx T \ln(\Lambda L)$ is log divergent. Also,
$\langle (\delta {\bf N})^2(T) \rangle$ is well behaved at all temperatures.   

In the presence of a substrate, the translational symmetry along the $z$ direction is broken and a
new term is allowed:
\begin{eqnarray}
U_S = \frac{v}{2} \int d^2 r (h({\bf r})-s({\bf r}))^2 \, ,
\label{substrate}
\end{eqnarray}
where $v$ is the interaction strength with the substrate and $s({\bf r})$ is a reference height.
Notice that in this case the flexural mode becomes gapped (by means of the $h^2$ term) and hence
all fluctuations (quantum and thermal) are quenched. Therefore, the most generic energy for
small and smooth height distortions of graphene is:
\begin{eqnarray}
U = \frac{1}{2} \int d^2r \left\{\kappa \left[\nabla^2 h({\bf r})\right]^2 + \gamma
\left[\nabla h({\bf r})\right]^2 + v (h({\bf r})-s({\bf r}))^2 \right\} \, .
\label{surf}
\end{eqnarray} 

Let us consider the case of (\ref{surf}) in the presence of a substrate \cite{swain}.
Minimization of the free energy with respect to $h$ leads to the equation:
\begin{eqnarray}
\kappa \nabla^4 h - \gamma \nabla^2 h + v h = v s \, .
\label{minh}
\end{eqnarray}
A particular solution of the non-homogeneous equation can be obtained
by Fourier transform:
\begin{eqnarray}
h({\bf q}) = \frac{s({\bf q})}{1+\ell_t^2 q^2 + \ell_c^4 q^4} \, ,
\label{hz}
\end{eqnarray}
where
\begin{eqnarray}
\ell_t &=& \left(\frac{\gamma}{v}\right)^{1/2} \, ,
\nonumber
\\
\ell_c &=& \left(\frac{\kappa}{v}\right)^{1/4} \, ,
\end{eqnarray}
are the length scales associated and curvature, respectively. 
Notice that (\ref{hz}) implies that the surface height more or less
follows the substrate landscape, as expected. 

\begin{figure}[tbh]
\centerline{\includegraphics[width=6cm, keepaspectratio]{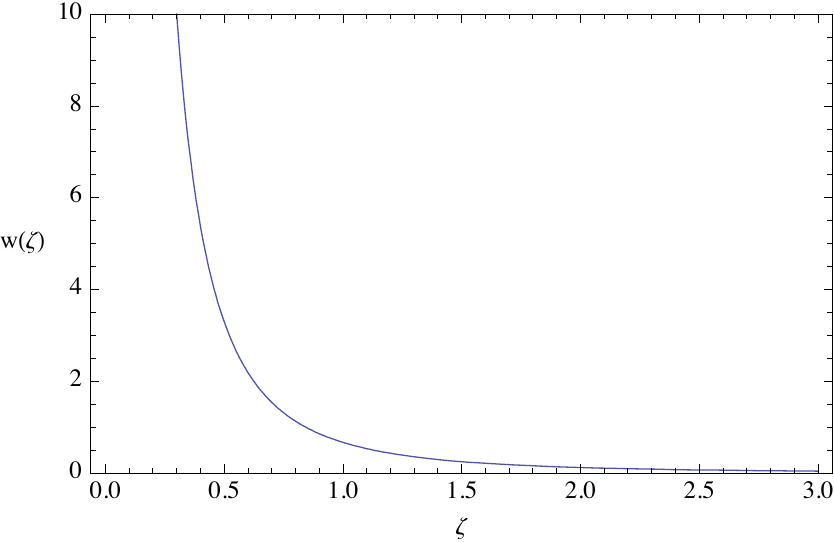}}
\caption{Amplitude of the local height variations in graphene supported in a random substrate, $w(\zeta)$.}
\label{wzeta}
\end{figure}

Let us consider the case of a random substrate where the probability of
a substrate height between $s$ and $s+ds$ is given by:
\begin{eqnarray}
P(s) = \frac{1}{\cal{N}} \exp\left\{-\int d^2r \frac{s^2}{2 s_0^4}\right\} \, ,
\end{eqnarray}
where ${\cal N}$ is a normalization factor:
\begin{eqnarray}
{\cal N} = \int_{-\infty}^{+\infty} ds P(s) = \left(\frac{\sqrt{2 \pi} s_0^2}{a}\right)^N \, ,
\end{eqnarray}
where $N$ is the number of surface cells, $a$ is their lattice spacing, and $s_0$ is the average height variation. In this case, the height correlation function can be shown to be:
\begin{eqnarray}
\overline{s({\bf r}) s({\bf r'})} = s_0^4 \delta({\bf r}-{\bf r'}) \, .
\label{stat}
\end{eqnarray}
In momentum space (\ref{stat}) is written as:
\begin{eqnarray}
\overline{s({\bf q}) s({\bf q'})} = (2 \pi)^2 s_0^4 \delta({\bf q}+{\bf q'}) \, .
\label{statk}
\end{eqnarray}
Using  (\ref{statk}) we can immediately compute the height-height correlation function as a function of the in-plane distance: 
\begin{eqnarray}
 \overline{h(r) h(0)}&=& \int \frac{d^2q}{(2\pi)^2} \int \frac{d^2q'}{(2\pi)^2} 
\frac{e^{i {\bf q} \cdot {\bf r}}}{(1+\ell^2_t q^2 + \ell^4_c q^4)(1+\ell^2_t (q')^2+\ell^4_c (q')^4)}
\overline{s({\bf q}) s({\bf q'})}
\nonumber
\\
&=& \frac{s_0^4}{2 \pi} \int_0^{\infty} dq \frac{q J_0(q r)}{(1+\ell^2_t q^2 + \ell^4_c q^4)^2}
= \frac{s_0^4}{2 \pi \ell^8_t} \int_0^{\infty} dq \frac{q J_0(q r)}{(q^2+P_+^2)^2(q^2+P_-^2)^2} \, ,
\end{eqnarray}
where we have defined:
\begin{eqnarray}
P^2_{\pm} &=& \frac{1}{2 \ell^2} \left(1 \pm \sqrt{1-\zeta^2}\right) \, ,
\nonumber
\\
\ell &=& \frac{\ell_c^2}{\ell_t} = \sqrt{\frac{\kappa}{\gamma}} \, ,
\nonumber
\\
\zeta &=& 2 \left(\frac{\ell_c}{\ell_t}\right)^2 = 2 \sqrt{\frac{\kappa v}{\gamma^2}} \, ,
\end{eqnarray}
and the behavior of the integral depends on whether $\zeta$ is smaller or bigger than $1$. For $r \gg \lambda = 1/{\rm Min}({\rm Re}(P_\pm))$ one finds:
\begin{eqnarray}
\overline{h(r) h(0)} \sim \frac{s_0^4}{2 \pi \lambda^2} e^{-r/\lambda} \, ,
\end{eqnarray}
showing that the height fluctuations are short ranged and decay with a characteristic length scale given by $\lambda$. 
The local height variations are given by:
\begin{eqnarray}
 \overline{h^2(0)}&=& \frac{s_0^4}{2 \pi \ell^2_t} w_{\pm}(\zeta) \, ,
\end{eqnarray}
where for $\zeta>1$ we have:
\begin{eqnarray}
w_+(\zeta) = \frac{1}{(\zeta^2-1)^{3/2}} 
\left[\frac{\pi}{2} - \arctan\left(\frac{1}{\sqrt{\zeta^2-1}}\right) - \frac{\sqrt{\zeta^2-1}}{\zeta^2}\right] \, ,
\end{eqnarray}
and for $\zeta<1$:
\begin{eqnarray}
 w_-(\zeta) = \frac{1}{(1-\zeta^2)^{3/2}} \left[\frac{\sqrt{1-\zeta^2}}{\zeta^2} - 
\frac{1}{2} \ln\left(\frac{1+\sqrt{1-\zeta^2}}{1-\sqrt{1-\zeta^2}}\right)\right] \, ,
\end{eqnarray}
the function $w(\zeta)$ is shown in Fig. \ref{wzeta}. Notice as $\zeta \gg 1$, that is, when $\ell_c \gg \ell_t$, the height variations are strongly suppressed and the system is essentially flat. This happens because in this limit the interaction with the substrate is strong, and the bending rigidity is large, compared with the tension.

\section{Deformed Graphene}
\label{deformed}

Graphene is a one atom thick and hence very soft. Just as any soft material, graphene can be easy deformed, especially in the out of the plane direction where the restitution forces vanish.
These deformations couple directly to the electrons, making of graphene a unique example of a metallic membrane \cite{kim}. 
Let us consider what happens to the electrons when the graphene is deformed either by bending or strain.
One effect is the change in the distance between the atoms, the other is the change in the overlap between the different orbitals. 
In both cases, the hopping energy between different carbon atoms is affected.  Consider the case that of the nearest neighbor hopping that changes from $t_0$ 
to $t_0 + \delta t_i(\vec{\delta})$ at site ${\bf R}_i$ in the
direction of $\vec{\delta}$. In this case we have to add a new term to Hamiltonian (\ref{basichamil}):
\begin{eqnarray}
\delta H_0 &=& - \sum_{{\bf k},{\bf k'}} a^{\dag}_{{\bf k}} b_{{\bf k'}}
\sum_{i,\vec{\delta}} \delta t_i(\vec{\delta}) e^{i({\bf k}-{\bf k'}) \cdot
  {\bf R}_i - i \vec{\delta} \cdot {\bf k'}}
\nonumber
\\
&\approx& - \sum_{{\bf q},{\bf q'}} \psi^{\dag}_{A,{\bf q}} \psi_{B,{\bf q'}}
\sum_{i,\vec{\delta}} \delta t_i(\vec{\delta}) e^{i ({\bf q}-{\bf q'}) \cdot
  {\bf R}_i} e^{-i \vec{\delta} \cdot {\bf Q}_1}  
\nonumber
\\
&\approx& \int d^2 r \, \, \psi^{\dag}_A({\bf r}) \psi_{B}({\bf r}) 
{\cal A}({\bf r}) \, , 
\label{dha}
\end{eqnarray}
where
\begin{eqnarray}
{\cal A}({\bf r}) = {\cal A}_x({\bf r}) + i {\cal A}_y({\bf r}) = 
- \sum_{\vec{\delta}} \delta t_i(\vec{\delta}) 
e^{-i \vec{\delta} \cdot {\bf Q}_1} \, .
\label{gaugefield}
\end{eqnarray}

Notice that if we replace (\ref{dha}) into (\ref{dirac}) the
full Hamiltonian becomes:
\begin{eqnarray}
H = \int d^2 r \Psi^{\dag}({\bf r}) \left[\vec{\sigma} \cdot \left(i v_F 
\nabla + \vec{{\cal A}}\right) - \mu \right] \Psi({\bf r}) \, ,
\label{full1}
\end{eqnarray}
where $\vec{{\cal A}} = ({\cal A}_x,{\cal A}_y)$ and hence the changes in
the nearest neighbor hopping couple like a gauge field. One should not worry about a possible broken time reversal symmetry here because
there is another nonequivalent point in the Brillouin zone, at K', which is related to the K point by inversion symmetry. Hence,
in the K' point the ``magnetic field'' is reversed and therefore there is no net time reversal symmetry breaking. 

Let us consider the problem of the hopping between next nearest
neighbor sites. The Hamiltonian in this case is:
\begin{eqnarray}
H_1 = - t^{'}_0 \sum_{\langle i, j\rangle} \left(a^{\dag}_i a_j + b^{\dag}_i b_j
  + {\rm h.c.} \right)
\label{h1}
\end{eqnarray}
which describes essentially hopping on a triangular lattice with lattice spacing $\sqrt{3} a$ with nearest
neighbor vectors:
\begin{eqnarray}
\vec{v_1} &=& \sqrt{3} a (0,1) \, ,
\nonumber
\\
\vec{v_2} &=& \sqrt{3} a (\sqrt{3}/2,-1/2) \, ,
\nonumber
\\
\vec{v_3} &=& \sqrt{3} a (-\sqrt{3}/2,-1/2) \, .
\label{nnvt}
\end{eqnarray} 

By Fourier transforming (\ref{h1}) we find:
\begin{eqnarray}
H_1 = \sum_{{\bf k}} \epsilon({\bf k}) 
\left(a^{\dag}_{\bf k} a_{\bf k} + b^{\dag}_{\bf k} b_{\bf k} \right) \, ,
\label{h1f}
\end{eqnarray}
where
\begin{eqnarray}
\epsilon({\bf k}) &=& -t^{'}_0 \sum_{\vec{v}} e^{i {\bf k} \cdot \vec{v}}
\nonumber
\\
&=& 2 t^{'}_0 \left[
\cos(\sqrt{3} a k_y) + 4 \cos(3 a k_x/2) \cos(\sqrt{3} a k_y/2)
\right] \, .
\end{eqnarray}
For ${\bf k} = {\bf Q_1} + {\bf q}$ we find:
\begin{eqnarray}
\epsilon({\bf Q}_1 + {\bf q}) \approx - 3 t^{'}_0 + \frac{9 t^{'}_0 a^2}{2} q^2
\label{exptri}
\end{eqnarray}
for $q \ll Q_1$. Notice that the first term  (\ref{exptri}) 
leads to a shift in the chemical potential and the second term 
introduces a quadratic term in the dispersion. For $q \ll t_0/(3 a t^{'}_0)$
we can disregard this term and consider only the chemical potential shift.
{\it Ab initio} calculations estimate that $t^{'}_0 \approx 0.1 t_0$
providing a good range where this approximation is valid \cite{rmp09}. 

Consider again the deformations of the graphene surface. 
In this case $t^{'}_0$ in (\ref{exptri}) has to be replaced by
$t' = t^{'}_0 + \delta t'$ and, in complete analogy with
(\ref{dha}), we find:
\begin{eqnarray}
\delta H_1 \approx \int d^2 r \, \, \Phi({\bf r}) 
\left(\psi^{\dag}_A({\bf r}) \psi_{A}({\bf r}) + 
\psi^{\dag}_B({\bf r}) \psi_{B}({\bf r}) \right) \, ,
\end{eqnarray}
where
\begin{eqnarray}
\Phi({\bf r}) &=& - 3 \sum_{\vec{v}} \delta t^{'}_i(\vec{v}) 
e^{-i \vec{v} \cdot {\bf Q}_1} 
\label{shift}
\end{eqnarray}
where we have used (\ref{deltat}), (\ref{vecQ}), and (\ref{nnvt}).
In this case, changes in the next-to-nearest neighbor hopping couple
to the Dirac fermions as a scalar potential. The final Dirac Hamiltonian 
has the form:
\begin{eqnarray}
H = \int d^2 r \Psi^{\dag}({\bf r}) \left\{\vec{\sigma} \cdot \left(i v_F \nabla
+ \vec{{\cal A}}\right) -\mu + \Phi({\bf r}) \right\} \Psi({\bf r}) \, .
\label{diracfinal}
\end{eqnarray}
Given the local changes in the hopping energies the scalar and vector
potentials can be readily computed through (\ref{gaugefield}) and (\ref{shift}), respectively.

This changes in hopping energies can be connected to changes in the structure of the lattice. Let us consider the case of in-plane distortions. In this case, the only change in the hopping energy is due to the change in the distance between the $p_z$ orbitals. Consider the case where the distance between sublattices changes by a distance $\delta \ell$ in the direction of $\vec{\delta}$, 
in first order we have:
\begin{eqnarray}
\delta t \approx (\partial t/\partial a) \delta \ell \, ,
\\
\delta \ell \approx (\vec{\delta} \cdot \nabla) {\bf u} \, ,
\end{eqnarray}
where $u({\bf r})$ is the lattice displacement. Replacing the
above expression in (\ref{gaugefield}) we get:
\begin{eqnarray}
{\cal A}^{(u)}_x({\bf r}) &\approx& \alpha a \, (u_{xx}-u_{yy}) \, ,
\nonumber
\\
{\cal A}^{(u)}_y({\bf r}) &\approx& \alpha a \, u_{xy} \, ,
\label{gaugein}
\end{eqnarray}
where $\alpha$ is a constant with dimensions of energy and we have used the standard definition of the strain tensor:
\begin{eqnarray}
u_{ij} = \frac{1}{2} \left(\frac{\partial u_i}{\partial x_j}+ \frac{\partial u_j}{\partial x_i}\right) \, .
\label{strain}
\end{eqnarray}
An analogous calculation for the change in the next-to-nearest neighbor hopping energy leads to:
\begin{eqnarray}
\Phi^{(u)}({\bf r}) \approx g (u_{xx}+u_{yy}) \, .
\label{potin}
\end{eqnarray}
Equations (\ref{gaugein}) and (\ref{potin}) relate the strain tensor to potentials that couple directly to the Dirac particles.

\begin{figure}[tbh]
\centerline{\includegraphics[width=8cm, keepaspectratio]{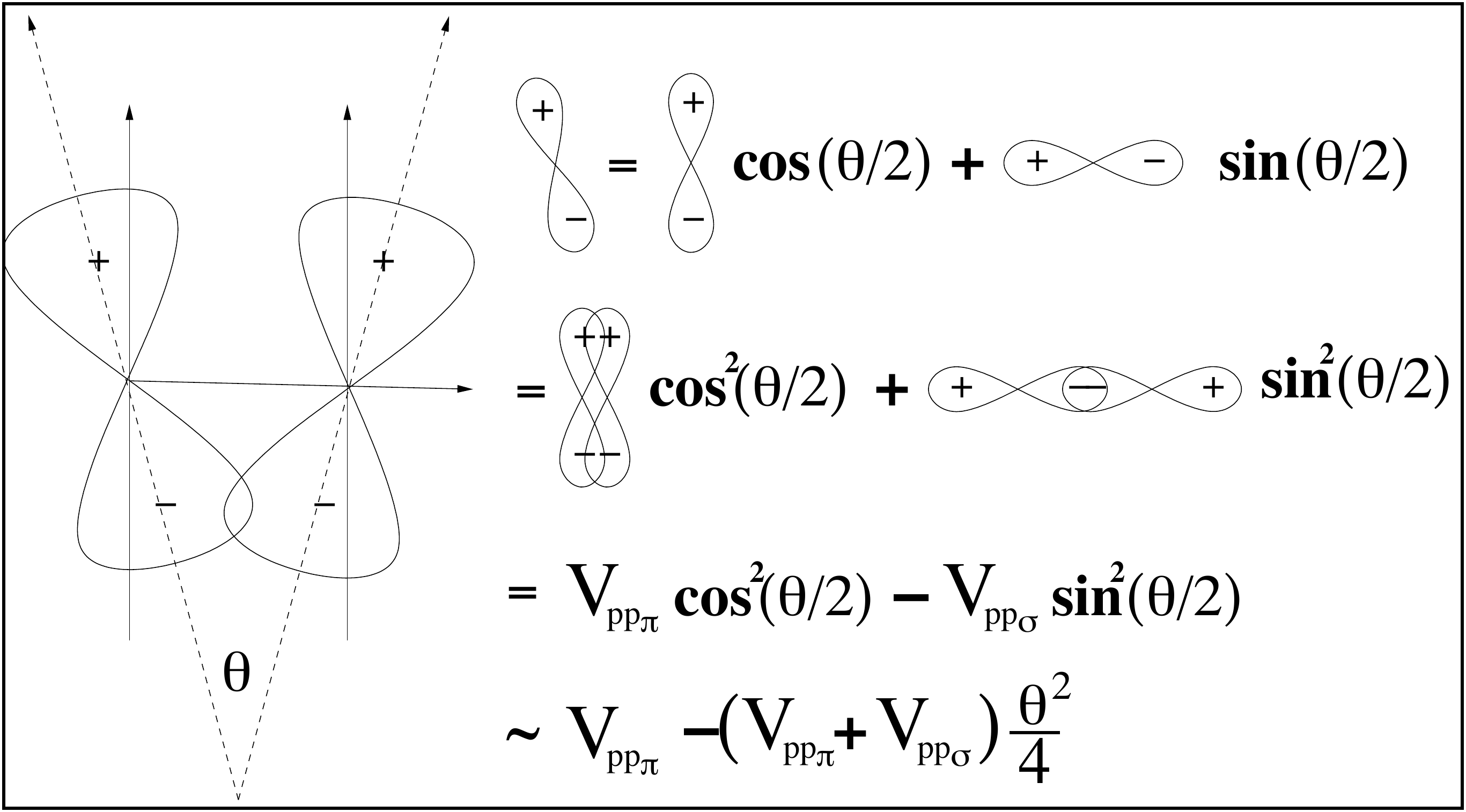}}
\caption{Rotation of p$_z$ orbitals.}
\label{rotation}
\end{figure}

Less trivial is the coupling to the out-of-plane modes since those involve rotations of the orbitals. In Fig.\ref{rotation} we show a rotation of two orbitals by an angle $\theta$. The rotation mixes $\pi$ and $\sigma$ states and for small angles the change in the hybridization energy is given by:
\begin{eqnarray}
t \approx t_0 + \delta V \theta^2 \, ,
\end{eqnarray}
where $\delta V$ is the energy mixing between $\pi$ and $\sigma$ states. Notice that $\theta =  2 \pi/R$ where $R$ is the curvature radius. In terms of the height variable we can write, in analogy with (\ref{over}), that the change in the hopping amplitude due to bending in the direction ${\bf u}$ is given by:
\begin{eqnarray}
\delta t \approx \delta V  \left[({\bf u} \cdot \nabla) \nabla h\right]^2 \, .
\end{eqnarray}
On the one hand, if ${\bf u}$ is a nearest neighbor vector $\vec{\delta}$ we get from (\ref{gaugefield}):
\begin{eqnarray}
{\cal A}^{(h)}_x({\bf r}) &= &- \frac{3 E_{ab} a^2}{8} \left[ (
\partial_x^2 h )^2 -
( \partial_y^2 h )^2 \right]  \, ,
\nonumber \\
{\cal A}^{(h)}_y({\bf r}) &= &\frac{3 E_{ab} a^2}{4} \left( \partial_x^2 h
+ \partial_y^2 h \right) \partial_x h \partial_y h \, ,
\label{field_curvature}
\end{eqnarray}
where the coupling constant $E_{ab}$ depends on microscopic details. On the other hand, if ${\bf u}$ is the next-to-nearest neighbor vector we find, in accordance with (\ref{shift}):
\begin{eqnarray}
\Phi^{(h)}({\bf r}) \approx - E_{aa} a^2 \left[\nabla^2 h({\bf r})\right]^2 \, ,
\end{eqnarray}
where $E_{aa}$ is an energy scale associated with the mixing between orbitals. The main conclusion is therefore that for smooth distortions of graphene due to strain or bending the
Dirac particles are subject to scalar and vector potentials leading to an ``electrodynamics'' that is purely geometrical (there is no electric charge associated with the ``electric'' and
``magnetic'' fields created by structural deformations). This structural ``electrodynamics'' has strong consequences for the electronic motion in graphene leading to many unusual effects
that cannot be found in ordinary materials. In particular, one can manipulate the electrons by ``constructing'' appropriate deformations of the lattice that mimic electric and magnetic
fields. This is the so-called {\it strain engineering} and is a field of research that is still in its infancy \cite{vitor,paco}. 

\subsection{A non-trivial example: The Scroll \cite{new}}
\label{scroll}
 
So far we have discussed small distortions of the graphene sheet, but as a soft material graphene can be bent by large angles making completely new structures. 
The scroll is an example of the softness of graphene that shows up as an interplay between the van der Waals energy, $E_{vdW}$, that makes graphene stick to itself and the bending energy, 
$E_B$, the energy cost to introduce curvature in the graphene sheet. Let us show first that the scroll is a stable configuration of a free standing graphene sheet. 

\begin{figure}[tbh]
\centerline{\includegraphics[width=8cm, keepaspectratio]{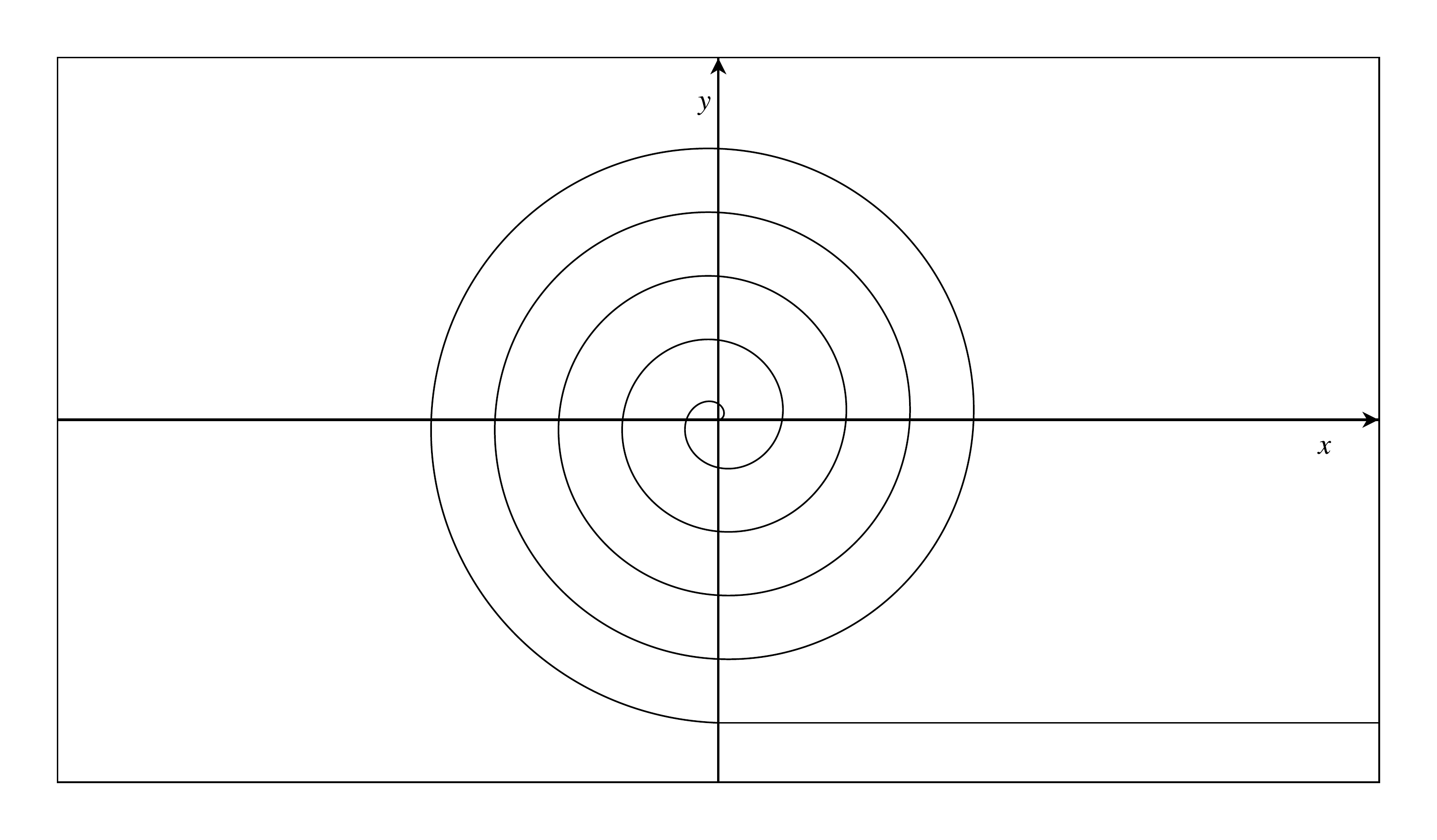}}
\caption{Graphene scroll in the form of an Archemedian spiral.}
\label{spiral}
\end{figure}

\begin{figure}[tbh]
\centerline{\includegraphics[width=8cm, keepaspectratio]{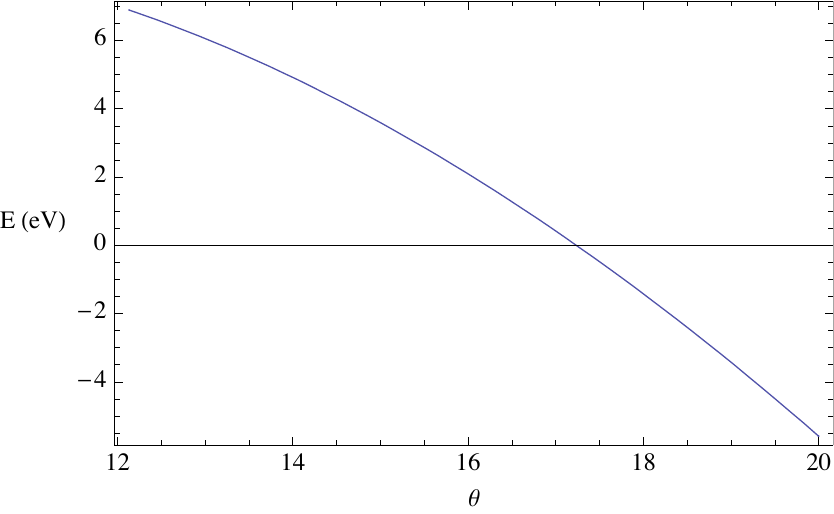}}
\caption{Energy of a scroll as a function of angle of rotation.}
\label{scrollene}
\end{figure}

We assume here that when graphene forms a scroll the equilibrium distance between folded regions is the same that one would find between the planes in graphite, that is, $d = 0.34$ nm. 
In this case, the scroll is described by the so-called Archemedian spiral:
\begin{eqnarray}
r(\theta) = a \theta \, ,
\label{spir}
\end{eqnarray}
where $a = d/(2\pi)$. We assume that the scroll has length $L$ along the scroll axis and consider the two main energy scales in this problem:
\begin{eqnarray}
\frac{E_{vdW}}{L} &=& - \gamma \int_{\theta_0+2 \pi}^{\theta_1} ds(\theta) \, ,
\label{vdw}
\\
\frac{E_B}{L} &=& \frac{\kappa}{2} \int_{\theta_0}^{\theta_1} ds(\theta) \frac{1}{R^2(\theta)}  \, ,
\label{ebend}
\end{eqnarray}
where $\gamma \approx 2.5$ ev/nm is the van der Waals coupling \cite{tomasek}, $\kappa \approx 1$ eV is the bending energy \cite{rmp09}. The radius of curvature of
the Archemedian spiral is given by:
\begin{eqnarray}
R(\theta) = \frac{a (1+\theta^2)^{3/2}}{2+\theta^2} \, ,
\label{radius}
\end{eqnarray}
and 
\begin{eqnarray}
ds = a \sqrt{1+\theta^2} d\theta \, ,
\label{ds}
\end{eqnarray}
is the infinitesimal arc-length of the spiral. The evaluation of the integrals is straightforward:
 \begin{eqnarray}
\frac{E_{vdW}}{L} &=& - \frac{\gamma a}{2} \left[\ln\left(\frac{\theta_1+\sqrt{1+\theta_1^2}}{(\theta_0+2 \pi)+\sqrt{1+(2 \pi+\theta_0)^2}}\right)+\frac{\theta_1 (9+8\theta_1^2)}{(1+\theta_1^2)^{3/2}}-\frac{(\theta_0+ 2\pi) (9+8(\theta_0+2 \pi)^2)}{(1+(\theta_0+2\pi)^2)^{3/2}}\right] \, ,
\label{vdwteta}
\\
\frac{E_B}{L} &=& \frac{\kappa a}{2} \left[
\ln\left(\frac{\theta_1+\sqrt{1+\theta_1^2}}{\theta_0+\sqrt{1+\theta_0^2}}
\right)+\theta_1 (1+\theta_1^2)^{1/2}-\theta_0 (1+\theta_0^2)^{1/2}\right] \, .
\label{ebendteta}
\end{eqnarray}
The angle $\theta_0$ is determined by the condition that the force on the graphene sheet vanishes at the edge of the scroll:
\begin{eqnarray}
F = - \frac{1}{L} \frac{\partial E}{\partial \theta_0}= - \frac{\kappa}{2 a}
\frac{(2+\theta_0^2)^2}{(1+\theta_0^2)^{5/2}} + \gamma a \sqrt{1+(\theta_0+2 \pi)^2} = 0 \, ,
\end{eqnarray}
and using the values of the parameters we find $\theta_0 \approx 5.85$. The total energy as a function of $\theta_1$ is given by (\ref{vdwteta}) and (\ref{ebendteta}) and shown in Fig.\ref{scrollene}. Notice that the scroll becomes stable for angles of rotation bigger than $\approx 17.23$ which is equivalent to approximately $3$ turns of the scroll. This result is in agreement with more involved calculations \cite{douglas}.

\begin{figure}[tbh]
\centerline{\includegraphics[width=8cm, keepaspectratio]{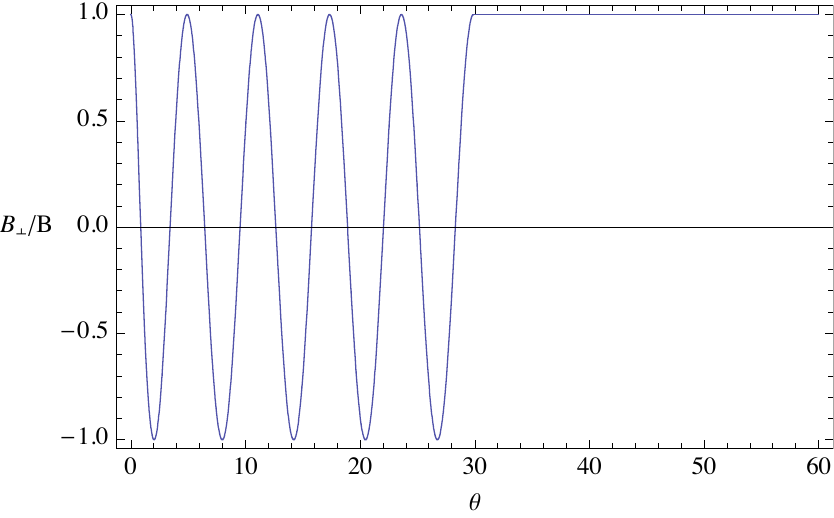}}
\caption{Magnetic field perpendicular to the graphene close to the scroll.}
\label{scrollmag}
\end{figure}

Consider what happens when a {\it uniform} magnetic field is applied perpendicular to the graphene plane. Notice that in the region of the scroll the component of the field perpendicular to the graphene surface is changing periodically. In fact, the normal component to graphene sheet at the scroll can be written as:
\begin{eqnarray}
{\bf N}(\theta) = \frac{1}{\sqrt{1+\theta^2}} (-\sin(\theta)-\theta \cos(\theta),\cos(\theta)-\theta \sin(\theta)) \, ,
\end{eqnarray}
and hence the component of the field perpendicular to the graphene is:
\begin{eqnarray}
B_{\perp}(\theta) = {\bf B} \cdot {\bf N}(\theta) = \frac{B}{\sqrt{1+\theta^2}} (\cos(\theta)-\theta \sin(\theta)) \, ,
\end{eqnarray}
which is shown in Fig. \ref{scrollmag}. Notice that from the point of view of the electrons, which are constrained to live in a two dimensional universe, a uniform magnetic field in three dimensions, becomes an oscillating magnetic
field in the presence of the scroll. An oscillating magnetic field has unusual effects on the electronic motion because the Lorentz force changes sign in the region where the field changes sign
producing a ``snake-like'' motion around the regions of zero field. These {\it snake states} are chiral one-dimensional correlated systems with rather unusual properties \cite{snake}. While it is very hard to produce a magnetic field that oscillates in a short length scale, it is very hard to roll graphene on a short length scale. Therefore, snake states should be easy to observe in rolled graphene sheets.

 \section{Conclusions}
\label{conclusions}

Graphene is a unusual system that shares properties of soft and hard matter and mimics problems in particle physics. These properties have their origin on the nature of its chemical bonds, the $s-p$ hybridization, and the low dimensionality. All these ingredients create a new framework for theoretical exploration which is still very much in its beginning. The possibility of studying electronics of deformed surfaces is quite intriguing because of its exotic ``electrodynamics''. The field of electronic membranes was born out of the discovery of graphene in 2004 and much has still to be understood.

In these notes I covered some very basic aspects of graphene chemistry and physics. We have studied the basic Hamiltonian that describes the overall band structure of graphene. We have seen that flexural modes, nonexistent in three dimensional solids, are fundamental for the understanding of the structural stability of the material. We have seen that substrates can change considerably the flexural modes and hence control the height fluctuations in supported samples. Finally, we have studied the various ways that graphene deformations, either by strain or bending, can modify the electron propagation in this material. In particular, new structures that can be created out of flat graphene, such as the graphene scroll, can have exotic properties, such as snake states, in the presence of applied uniform magnetic fields. In some way, the present notes complete some of the material already published in ref. \onlinecite{rmp09}. Nevertheless, the whole subject of graphene electronics and structure is much bigger than that and there is so much more to be understood.

It is quite obvious that a material that is structurally robust, still flexible, and extremely clean, has enormous potential for technological applications. By understanding these unusual properties one can harvest new functionalities that did not exist before. Progress in material science unavoidably leads to progress. Graphene is considered today one of the most promising candidates for a new era of a carbon-based technology, possibly supplanting the current silicon-based one. However, only progress in the understanding the basic properties of these systems can actually drive the technological progress. 

{\it Acknowledgements}. It is a pleasure to acknowledge countless conversations with Eva Andrei, Misha Fogler, Andre Geim, Francisco Guinea, Silvia Kusminskiy, Valeri Kotov, Alessandra Lanzara, Caio Lewenkopf, Johan Nilsson, Kostya Novoselov, Eduardo Mucciolo, Vitor Pereira, Nuno Peres, Marcos Pimenta, Tatiana Rappoport, Jo\~ao Lopes dos Santos, and Bruno Uchoa. This work was possible due to the financial support of a Department of Energy grant DE-FG02-08ER46512 and the Office of Naval research grant MURI N00014-09-1-1063.


\begin{thebibliography}{100}

\bibitem{rmp09} A. H. Castro Neto, F. Guinea, N. M. R. Peres, K. S. Novoselov, and A. K. Geim, {\it The Electronic Properties of Graphene}, Reviews of Modern Physics {\bf 81}, 109 (2009).

\bibitem{novo04} K. S. Novoselov, A. K. Geim, S. V. Morozov, D. Jiang, Y. Zhang, S. V. Dubonos, I. V. Grigorieva, A. A. Firsov, {\it Electric Field Effect in Atomically Thin Carbon Films}, Science {\bf 22}, 666 (2004).

\bibitem{rise} A. K. Geim, and K. S. Novoselov, {\it The rise of graphene}, Nature Materials {\bf 6}, 183 (2007).

\bibitem{shankar}R. Shankar, {\it Renormalization Group Approach to Interacting Fermions}, Reviews of Modern Physics {\bf 66}, 129 (1994).

\bibitem{baym}G. Baym and C. Pethick, {\it Landau Fermi-Liquid Theory} (Wiley, New York, 1991). 

\bibitem{pw06} A. H. Castro Neto, F. Guinea, and N. M. R. Peres, {\it Drawing conclusions from graphene}, 
Physics World {\bf 19}, 33 (2006).

\bibitem{gonzalez}J. Gonzalez, F. Guinea, and M. A. H. Vozmediano, {\it Non-Fermi liquid behavior of electrons in the half-filled honeycomb lattice: a renormalization group approach}, Nuclear Physics B {\bf 424}, 595 (1994).

\bibitem{note1} It should be noted that, even if a gap can be opened in graphene (as if, for instance, one finds a way to break the symmetry between different sublattices), its dispersion would be hyperbolic, not parabolic, because the very basic Lorentz invariance is preserved.

\bibitem{baym2} G. Baym, and S. A. Chin, {\it Landau Theory of Relativistic Fermi Liquids}, Nuclear Physics A {\bf 262}, 527 (1976),

\bibitem{Baymbook} Gordon Baym, {\it Lectures on Quantum Mechanics} (Addison-Wesley, Reading, 1990).

\bibitem{Pauling}L. Pauling, {\it The Nature of the Chemical Bond} (Cornell University Press, Ithaca, 1960).

\bibitem{sographene} A. H. Castro Neto and F. Guinea, {\it Impurity induced spin-orbit coupling in graphene}, Physical Review Letters {\bf 103}, 026804 (2009).

\bibitem{harrison} W. A. Harrison, {\it Elementary Electronic Structure} (World Scientific, Singapore, 2005).

\bibitem{painter}G. S. Painter and D. E. Ellis, {\it Electronic Structure and Optical Properties of Graphite from a Variational Approach}, Physical Review B {\bf 1} 4747 (1970).

\bibitem{chaikin}P. Chaikin, and T. C. Lubensky, {\it Introduction to Condensed Matter Physics} (Cambridge University Press, Cambridge, 1995).

\bibitem{swain}P.S. Swain, and D. Andelman, {\it The influence of substrate structure on membrane adhesion}, Langmuir {\bf 15}, 8902 (1999).

\bibitem{kim} Eun-Ah Kim and A. H. Castro Neto, {\it Graphene as an electronic membrane}, Europhysics Letters {\bf 84}, 57007 (2008).

\bibitem{vitor}V. M. Pereira, and A. H. Castro Neto, {\it All-graphene integrated circuits via strain engineering}, Physical Review Letters 103, 046801 (2009).

\bibitem{paco}F. Guinea, M. I. Katsnelson, and A. K. Geim, {\it Energy gaps, topological insulator state and zero field quantum Hall effect in graphene by strain engineering}, 
Nature Physics {\bf 6}, 30 (2010).

\bibitem{new} M. M. Fogler, A. H. Castro Neto, and F. Guinea, {\it Effect of external conditions on the structure of scrolled graphene edges}, Phys. Rev. B {\bf 81}, 161408 (R) (2010).

\bibitem{tomasek}D. Tom\'anek, {\it Mesoscopic origami with graphite: scrolls, nanotubes, peapods}, Physica B {\bf 323}, 86 (2002).

\bibitem{douglas}S. F. Braga, V. R. Coluci, S. B. Legoas, R. Giro, D. S. Galv\~ao, and R. H. Baughman, {\it Structure and Dynamics of Carbon Nanoscrolls}, Nano Letters {\bf 4}, 881 (2004).

\bibitem{snake} J. E. M\"uller, {\it Effect of a nonuniform magnetic field on a two-dimensional electron gas in the ballistic regime}, Physical Review Letters {\bf 68}, 385 (1992).

\end{thebibliography}
\end{document}